\documentclass[journal]{IEEEtran}
\usepackage{subcaption}
\usepackage{amsmath, amssymb, bm, mathrsfs, xfrac, mathtools, amsthm}
\usepackage{lettrine}
\usepackage{pgf}
\usepackage{booktabs}
\usepackage{stfloats}
\usepackage{tabularx}
\usepackage{cite}
\usepackage{multirow}
\usepackage{hyperref}
\usepackage{pifont}
\usepackage{graphicx}
\newcommand\sbullet[1][.5]{\mathbin{\vcenter{\hbox{\scalebox{#1}{$\bullet$}}}}}

\newtheorem{remark}{Remark}

\ifCLASSINFOpdf
\else
\fi

\begin{document}
\bstctlcite{IEEEexample:BSTcontrol}
\title{Observability and Parameter Estimation of a Generic Model for Aggregated Distributed Energy Resources \\[1ex] 
{\small }} 
\author{Bukunmi G. Odunlami,~\IEEEmembership{Student Member,~IEEE}, Marcos Netto,~\IEEEmembership{Senior Member,~IEEE}
\thanks{This work is supported by the National Science Foundation under Grant 2328241. The authors are with the Department of Electrical and Computer Engineering, New Jersey Institute of Technology, Newark, NJ 07102, USA.}
}

\maketitle

\begin{abstract}
We propose a novel framework for estimating the parameters of an aggregated distributed energy resources (DER\_A) model. First, we introduce a rigorous method to determine whether all model parameters are estimable. When they are not, our approach identifies the subset of parameters that can be estimated. The proposed framework offers new insights into the number and specific parameters that can be reliably estimated based on commonly available measurements. It also highlights the limitations of calibrating such models. Second, we introduce a Kalman filtering method to calibrate the DER\_A model. Since we account for nonlinear effects such as saturation and deadbands, we develop a specific mechanism to handle smoothing functions within the Kalman filter. Specifically, we consider the extended and the unscented Kalman filter. We demonstrate the effectiveness of the proposed framework on a modified IEEE 34-node distribution feeder with inverter-based resources. Our findings align with the North American Electric Reliability Corporation's parameterization guideline and underscore the importance of model calibration in accurately capturing the collective dynamics of distributed energy resources installed on distribution systems.
\end{abstract}


\IEEEpeerreviewmaketitle

\section*{Nomenclature}
\addcontentsline{toc}{section}{Nomenclature}
\begin{IEEEdescription}[\IEEEusemathlabelsep\IEEEsetlabelwidth{$V_1,V_2,V_3$}]
{\color{black}
\item[$\text{dbd1, dbd2}$] Lower, upper voltage deadband threshold
\item[$\text{D}_{\text{dn}}, \text{D}_{\text{up}}$] Under-, over-frequency droop gain
\item[$\text{fbd1, fbd2}$] Lower, upper frequency deadband threshold
\item[$f_{\text{emin}},\, f_{\text{emax}}$] Minimum, maximum frequency error limits
\item[$\text{freq}$] Frequency
\item[$I_{\text{d}},\, I_{\text{q}}$] Direct, quadrature current
\item[$I_\text{max}$] Maximum converter current magnitude
\item[$I_{\text{pcmd}}$] Active power command
\item[$I_{\text{qcmd}}$] Reactive power command
\item[$I_{\text{qv}}$] Reactive current from the voltage error
\item[$k_{\text{qv}}$, $k_{\text{pg}}$, $k_{\text{ig}}$] Gain constants 
\item[$\ell,\, \mathsf{h}$] Lower, upper limit
\item[$m_{\text{v}}$] Voltage tripping logic output
\item[$P$, $Q$] Active, reactive power
\item[$P_{\text{a}}$] Anti-windup–corrected active power output
\item[$P_{\mathrm{lim}}$] Error input to the proportional–integral block
\item[$V$] Voltage magnitude
\item[$V_{\mathrm{frac}}$] Fraction of DER population that recovers when $V_{\ell{1}}\! <\! V\! <\! V_{\mathsf{h}1}$
\item[$V_{\mathsf{h}0},\, V_{\mathsf{h}1}$] High-voltage cut-out breakpoints
\item[$V_{\ell{0}},\, V_{\ell{1}}$] Low-voltage cut-out breakpoints
\item[$V_{\text{min}},\, V_{\text{max}}$] User-defined reconnection bounds
\item[$\text{SDBF}$] Smooth deadband function
\item[$\text{SSF}$] Smooth saturation function
\item [$T_{\text{g}}$] Current control time constant
\item [$T_{\text{iq}}$] Reactive power control time constant
\item [$T_{\text{p}}$] Active power time constant
\item [$T_{\text{pord}}$] Power order time constant
\item [$T_{\text{rf}}$] Frequency measurement time constant
\item [$T_{\text{rv}}$] Voltage measurement time constant
\item [$T_{\text{v}}$] Voltage cut-out time constant
\item[$t_{\ell{0}},\, t_{\ell{1}}$] Timers associated with $V_{\ell{0}}$ and $V_{\ell{1}}$
\item[$t_{\mathsf{h}0},\, t_{\mathsf{h}1}$] Timers associated with $V_{\mathsf{h}0}$ and $V_{\mathsf{h}1}$
\item[$\Delta f'$] Frequency error bounded by deadbands
\item[$\Delta P_{\mathrm{droop}}$] Power adjustment from droop control
\item[$\Delta t$] Sampling time of the input signal
\item[$(\cdot)_{\text{ref}}$] Reference signals}
\end{IEEEdescription}

\section{Introduction}
\IEEEPARstart{T}{he} increasing adoption of distributed energy resources (DERs) is transforming the dynamics of power systems. These resources are no longer just peripheral actors at the edge of the grid, but now represent a significant portion of net electricity generation in many areas. As of July 2025, the United States had installed over 39 gigawatts of behind-the-meter, small-scale residential solar capacity \cite{EIA2025} across more than 4 million systems nationwide. In many urban power distribution networks, DERs account for {\color{black}more than 20\%} of local generation capacity \cite{NREL2023}.

The response of distribution networks with significant DER capacity to transmission network faults affects overall power system dynamic performance \cite{Alvarez-Fernandez2020}. {\color{black}Therefore, the IEEE 1547-2018 \cite{IEEEStd1547-2018} specifies the technical requirements for the performance, operation, testing, and safety of the interconnection and interoperability of DERs with electric power systems. For instance, it mandates that DERs provide specific grid-support functions, such as voltage regulation and disturbance ride-through.} Power system stability models must accurately reflect the dynamic behavior of DERs. To that end, power system engineers rely on aggregated dynamic models guided by computational and practical considerations{\color{black}, as reflected in transmission planning studies conducted by utilities and system operators that employ aggregated DER models for bulk power system simulations \cite{Droge_CIGRE_2019, Bhattarai2019DER}}. Modeling each DER individually is unrealistic for large-scale simulations, and such granularity is unnecessary for bulk power systems analysis \cite{EPRI2019}. {\color{black} Indeed, the increasing complexity and scale of power systems require reduced-order models that capture essential dynamics while remaining computationally feasible for stability studies \cite{SauerPaiChow2018}.} 

Since 2016, the Western Electricity Coordinating Council (WECC) has led the development of the \emph{aggregated distributed energy resources} (DER\_A) \cite{Pourbeik2019}; a time-domain, positive-sequence model designed to describe---if properly parameterized---the combined dynamics of many DERs connected throughout distribution feeders within a distribution system. However, its practical use in stability studies hinges on having a reliable, systematic method for tuning its parameters. 

Recognizing the importance of proper parameterization, the North American Electric Reliability Corporation (NERC) has published a reliability guideline \cite{NERC2026}. Section \ref{sec.IV} of this paper emphasizes that the parameter values suggested in \cite{NERC2026} are intended as guidelines and do not eliminate the need for model calibration. There is a need for a principled, automated method to assist power utilities and system operators in parameterizing and regularly recalibrating their models as more DERs connect to the grid. 

The DER\_A {\color{black}model} has a total of 10 state variables and 48 parameters \cite{EPRI2019}, although these numbers can be lower since the model is configurable. A fundamental question---one that we raise and answer---is whether all 48 parameters can be estimated. Section \ref{sec.III} of this paper shows that the answer is no. There are not enough measurements to estimate all 48 parameters. Often, there will be just enough information to estimate a subset of those parameters, as we will demonstrate.

\begin{figure*}[hb!]
\centering
\includegraphics[width=1\linewidth]{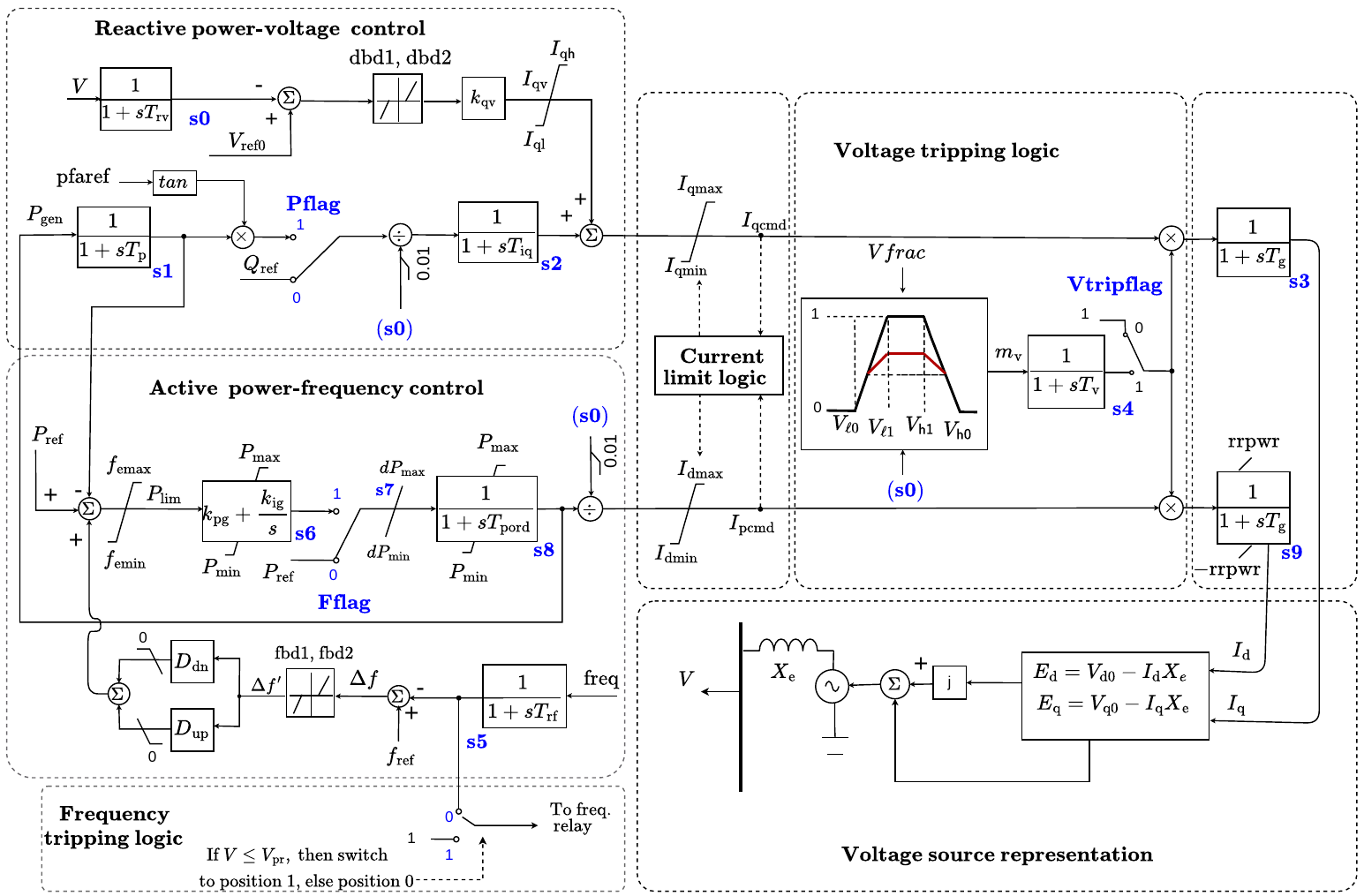}
\vspace{-.5cm}
\caption{{\color{black}Block diagram of the DER\_A model (adapted from \cite{NERC2026}), showing the control structure relevant to this study.}}
\label{fig:DER_A block diagram}
\end{figure*}

This challenge may have motivated \cite{Foroutan2023} to select control loops within the DER\_A {\color{black}model} and estimate their parameters, while assuming values for other parts of the model. Specifically, \cite{Foroutan2023} focuses on the parameters of the reactive power--voltage control loop (see Fig. \ref{fig:DER_A block diagram}). However, in their work \cite{Foroutan2023}, the authors remark that the estimated ``value of dbd1 was not reasonable." For the cases examined in this paper, we find that dbd1 and dbd2 are on the borderline of being estimable, thereby formally justifying the results in \cite{Foroutan2023}; see Table \ref{tab:EstimatedParams}. Although this conclusion depends on how one configures the DER\_A model and chooses the measurements, this paper offers a formal mechanism for verifying which parameters can be reliably estimated.

Since not all DER\_A model parameters can be estimated, as shown in Section \ref{sec.III}, the problem reduces to identifying the subset of parameters that can be reliably estimated. A few researchers have explored the DER\_A model parameterization by selecting a subset of perceived influential parameters. The work in \cite{9637939} proposes an ad hoc solution based on a nonlinear least squares optimization approach. It identifies, based on engineering judgment, a subset of the 10 most critical parameters. Then, it considers over 1,000 different combinations of those 10 parameters, and performs an exhaustive search to find the parameter combination that best fits (in a least squares sense) waveforms obtained from specific simulated disturbances.

{\color{black}In a related study, \cite{VasquezPlaza2024ParameterIF} considers parameter identification for the DER\_A voltage support function using a moving-horizon estimation framework formulated as a gradient-based optimization problem. The analysis is restricted to a fixed configuration of the reactive power–voltage control loop with a single control block, without examining the role of the control flag or justifying the selected configuration. As a result, the dependence of the input–output behavior and, consequently, the estimation performance on the control mode is not assessed.}

{\color{black}Similarly}, the work in \cite{MHE_Tonkoski} introduces an online moving-horizon estimation approach for the DER\_A {\color{black}model} parameterization, framing the problem as a constrained optimization over a receding time horizon. The study \cite{MHE_Tonkoski} suggests---based on computational simulations---that an extended Kalman filter (EKF) can reliably estimate up to three parameters. Conversely, this paper demonstrates that Kalman filter extensions designed for nonlinear systems are effective in estimating more than three parameters, provided the system is observable---an aspect not examined in \cite{MHE_Tonkoski}; see also \cite{4111008, EKF}. While moving horizon estimators are well-suited for constrained settings, their high computational demand hinders their real-time applicability. Kalman filtering, on the other hand, is widely used for state and parameter estimation \cite{Zhao2021}.



{\color{black}
Finally, we find that replicating earlier results \cite{Foroutan2023, 9637939, VasquezPlaza2024ParameterIF, MHE_Tonkoski} can be challenging. To address this, we have made our \texttt{MATLAB} code \cite{code} publicly available. We hope this paper will serve as a baseline for researchers, electric reliability organizations such as NERC and WECC, and, most importantly, electric utilities and system operators that use the DER\_A {\color{black}model} in their routine studies. The contributions of this paper are:

\begin{itemize}
\item \textbf{A rigorous observability-based framework for DER\_A parameter estimation.} We introduce a principled method to determine whether the parameters of the DER\_A model are estimable from available measurements. Unlike prior work that selected parameters heuristically, this framework formally identifies which parameters can and cannot be reliably estimated. 
\item \textbf{Identification of the estimable subset of DER\_A model parameters.} We show that not all parameters can be estimated because of insufficient measurement information and structural limitations in the model. By leveraging Lie-derivative observability analysis \cite{Hermann1977}, we determine the least observable parameters and obtain reduced parameter sets that become observable under practical measurement configurations.
\item \textbf{Development of a Kalman filtering calibration framework for nonlinear DER\_A models.} The paper develops both EKF and UKF approaches for online calibration of DER\_A models. The framework jointly estimates system states and parameters via an augmented-state formulation, enabling automated tuning of aggregated DER dynamics for power system studies.
\item \textbf{Novel treatment of nonlinear deadband and saturation functions for stable estimation.} The paper addresses a major challenge in DER\_A parameter estimation: discontinuous nonlinearities caused by saturation and deadband logic. It replaces these with smooth approximations and introduces a derivative regularization strategy that prevents Jacobian ill-conditioning and filter divergence. This contribution enables stable EKF/UKF estimation even for highly nonlinear inverter dynamics.
\end{itemize}

Our results show that the dynamics of the DER\_A model are more observable from power measurements than from current measurements, underscoring their importance in selecting measurements for model calibration.}

The paper proceeds as follows. Section \ref{sec.II} provides a brief overview of the DER\_A model. Section \ref{sec.III} introduces the proposed approach to observability and parameter estimation methods for the DER\_A model. Section \ref{sec.IV} presents numerical results, and Section \ref{sec.V} concludes the paper.

\section{The generic model for aggregated distributed energy resources}\label{sec.II}
This section provides a brief description of the DER\_A model. {\color{black}Fig. \ref{fig:DER_A block diagram} establishes a one-to-one correspondence between the DER\_A model and the parameter estimation formulation in this paper. The functional blocks are consistently labeled to enable a direct mapping between control loops and their parameters. Note that the DER\_A model includes variables (referred to as \emph{flags}) that enable/disable specific control and protection functions, making it adaptable to different distribution grids.} More specifically, these functions determine how DERs collectively respond to changes in voltage, frequency, and real and reactive power. There are two main control loops: reactive power--voltage and active power--frequency loops, and three control logics: voltage tripping, frequency tripping, and current limit. Note that the number of state variables and parameters in the DER\_A model varies depending on how the flags are configured. For simplicity, we denote by $x_{i}$ the $i$th state variable, and by $\kappa_{i}$ the $i$th flag, cf. Table \ref{tab:DER_A_StatesAndFlags}. Next, we elaborate on the \emph{current limit logic}, the \emph{voltage tripping logic}, and the \emph{deadband} and \emph{saturation} operators in Fig. \ref{fig:DER_A block diagram} {\color{black} to clarify how their configurations, through the flags, influence the input--output behavior and the resulting parameter estimation problem.}

The \emph{current limit logic} determines how DERs collectively respond as their output current approaches {\color{black}the maximum converter current,} $I_{\text{max}}$. This behavior is defined by $\kappa_{4}$, which specifies whether the inverter operates with active or reactive current priority, as shown in \eqref{eq:Qpriority}--\eqref{eq:Ppriority}. The priority logic sets the limits for {\color{black}the direct ($I_{\text{d}}$) and quadrature ($I_{\text{q}}$)} currents based on the selected mode and the total available current magnitude. When $\kappa_{4}=0$, the inverter operates in \textit{reactive power priority} mode, where \( I_{\text{q}} \) is maintained at its limit and \( I_{\text{d}} \) is adjusted accordingly. {\color{black}Formally,}
\vspace{-0.25cm}
\begin{align}
\text{reactive power priority} &=
\begin{cases} 
I_{\text{qmax}} = I_{\text{max}} \\
I_{\text{qmin}} = -I_{\text{max}} \\
I_{\text{dmax}} = \sqrt{I_{\text{max}}^2 - I_{\text{qcmd}}^2}
\end{cases}.
\label{eq:Qpriority} 
\end{align}

When $\kappa_{4}=1$, the inverter operates in \textit{active power priority} mode, maintaining \( I_{\text{d}} \) while \( I_{\text{q}} \) is modified to satisfy the current limit constraint. {\color{black}That is,}

\begin{align}
\text{active power priority} &=
\begin{cases}
I_{\text{dmax}} = I_{\text{max}} \\
I_{\text{qmax}} = \sqrt{I_{\text{max}}^2 - I_{\text{pcmd}}^2} \\
I_{\text{qmin}} = -I_{\text{qmax}}
\end{cases}.
\label{eq:Ppriority}
\end{align}

{\color{black}Note that a circular current capability curve is considered for both modes.} Following \cite{mathrep}, the expression for the \emph{voltage tripping logic}, denoted by $m_{\mathsf{v}}$, is as follows:
\begin{equation}\label{eq:Voltage logic}
m_{\mathsf{v}}=
\begin{cases}
\dfrac{V - V_{\ell{0}}}{V_{\ell{1}}-V_{\ell{0}}} & V_{\ell{0}} \le V \le V_{\text{min}} \\[8pt]
\dfrac{V - V_{\ell{0}}}{V_{\ell{1}}-V_{\ell{0}}} & V_{\text{min}} < V \le V_{\ell{1}},\ t \le t_{\ell{1}} \\[8pt]
1 & V_{\ell{1}} < V < V_{\mathsf{h}1}, \ t \le t_{\mathsf{h}1} \\
\dfrac{V_{\mathsf{h}0} - V}{V_{\mathsf{h}0}-V_{\mathsf{h}1}} & V_{\mathsf{h}1} \le V \le V_{\mathsf{h}0},\ t \le t_{\mathsf{h}1} \\[9pt]
V_{\mathrm{frac}} \dfrac{V - V_{\text{min}}}{V_{\ell{1}}-V_{\ell{0}}} & V_{\text{min}} \le V \le V_{\ell{1}},\ t \ge t_{\ell{1}} \\[8pt]
V_{\mathrm{frac}} \dfrac{V_{\ell{1}}-V_{\text{min}}}{V_{\ell{1}}-V_{\ell{0}}} & V_{\ell{1}} < V < V_{\mathsf{h}1},\ t \ge t_{\mathsf{h}1} \\[12pt]
V_{\mathrm{frac}} \dfrac{V_{\text{max}} - V}{V_{\mathsf{h}0}-V_{\mathsf{h}1}} & V_{\mathsf{h}1} \le V \le V_{\text{max}},\ t \ge t_{\mathsf{h}1} \\[8pt]
\dfrac{V_{\mathsf{h}0} - V}{V_{\mathsf{h}0}-V_{\mathsf{h}1}} & V_{\text{max}} < V \le V_{\mathsf{h}0} \\
0 & \text{otherwise.}
\end{cases}
\end{equation}
{\color{black}where $V_{\ell0}$, $V_{\ell1}$, $V_{\mathsf{h}0}$, $V_{\mathsf{h}1}$ are the voltage breakpoints defining the tripping characteristic.} Finally, we define the \emph{saturation} and \emph{deadband} operators as follows:
\begin{align}
[x]_{\ell}^{\mathsf{h}} &=
\begin{cases}
\mathsf{h} & x \geq \mathsf{h} \\
x & \ell < x < \mathsf{h} \\
\ell & x \leq \ell
\end{cases}, \label{eq:saturation} \\
\text{db}^{\Delta_2}_{\Delta_1}(x) &=
\begin{cases}
x - \Delta_2 & x > \Delta_2 \\
0 & \Delta_1 \leq x \leq \Delta_2 \\
x - \Delta_1 & x < \Delta_1
\end{cases}, \label{eq:deadband}
\end{align}

\begin{table}[!t]
\centering
\caption{{\color{black}The} DER\_A {\color{black}model} state variables and flags}
\label{tab:DER_A_StatesAndFlags}
\renewcommand{\arraystretch}{1.1}
\begin{tabular}{@{}l p{1cm} p{5.5cm}@{}}
\toprule
\textbf{State/Flag} & \textbf{Renamed} & \textbf{Physical meaning} \\
\midrule
$S_0$ & $x_1$ & Filtered terminal voltage \\
$S_1$ & $x_2$ & Filtered generated active power \\
$S_2$ & $x_3$ & Reactive current by Q or power factor control \\
$S_3$ & $x_4$ & Total reactive current injected by DERs \\
$S_4$ & $x_5$ & Trip voltage \\
$S_5$ & $x_6$ & Filtered frequency \\
$S_6$ & $x_7$ & Active power control effort \\
$S_7$ & $x_8$ & Ramp rate for active power output \\
$S_8$ & $x_9$ & Generated active power \\
$S_9$ & $x_{10}$ & Total active current injected by DERs \\
\midrule
Pflag & $\kappa_{1}$ & Reactive power control / constant reactive power \\
Fflag & $\kappa_{2}$ & Active power control / constant active power \\
Vtripflag & $\kappa_{3}$ & Voltage tripping logic \\
PQflag & $\kappa_{4}$ & Active/reactive power current priority \\
\bottomrule
\end{tabular}
\end{table}
{\color{black}\noindent where $\ell$ and $\mathsf{h}$ denote the lower and upper bounds}. With these definitions, we derive the ordinary differential and algebraic equations describing the dynamics of the DER\_A model:
\begin{align}
\dot{x}_1 &= -\frac{x_1}{T_{\text{rv}}} + \frac{V}{T_{\text{rv}}} \label{eq:x1_dynamics} \\
\dot{x}_2 &= -\frac{x_2}{T_{\text{p}}} + \frac{x_9}{T_{\text{p}}} \\
\dot{x}_3 &=
\begin{cases}
-\dfrac{x_3}{T_{\text{iq}}} + \dfrac{Q_{\text{ref}}}{T_{\text{iq}} \,[x_1]_{0.01}^{\infty}} & \text{if}\;\kappa_{1} = 0 \\[8pt]
-\dfrac{x_3}{T_{\text{iq}}} + \dfrac{x_2\,\tan(\mathrm{\text{pfaref}})}{T_{\text{iq}} \,[x_1]_{0.01}^{\infty}} & \text{otherwise}
\end{cases} \\
\dot{x}_4 &=
\begin{cases}
 -\dfrac{x_4}{T_{\text{g}}} + \dfrac{\left[\,x_3 - [I_{\text{qv}}]_{I_{\text{q$\ell$}}}^{I_{\text{q$\mathsf{h}$}}}\ \right]_{I_{\text{qmin}}}^{I_{\text{qmax}}}}{T_{\text{g}}} & \text{if}\;\kappa_{3} = 0 \\[6pt]
 -\dfrac{x_4}{T_{\text{g}}} + \dfrac{\left[\,x_3 - [I_{\text{qv}}]_{I_{\text{q$\ell$}}}^{I_{\text{q$\mathsf{h}$}}}\,\right]_{I_{\text{qmin}}}^{I_{\text{qmax}}} x_5}{T_{\text{g}}} & \text{otherwise}
\end{cases} \\
\dot{x}_5 &= -\frac{x_5}{T_{\text{v}}} + \frac{m_{\mathsf{v}}}{T_{\text{v}}} \\
\dot{x}_6 &= -\frac{x_6}{T_{\text{rf}}} + \frac{\text{freq}}{T_{\text{rf}}}\\ 
\dot{x}_7 &= k_{\text{ig}} P_{\text{lim}} + k_{\text{w}} \left( P_{\text{a}} - k_{\text{pg}} P_{\text{lim}} - x_7 \right) \\
\dot{x}_8 &=
\begin{cases}
0 & \text{if}\;\kappa_{2} = 0 \\
\left[ \dfrac{\left[ x_7 \right]_{P_\text{min}}^{P_\text{max}} - x_8}{\Delta t} \right]_{dP_\text{min}}^{dP_\text{max}} & \text{otherwise}
\end{cases} \\
\dot{x}_9 & = \frac{1}{T_{\text{pord}}}\, x_8 - \frac{1}{T_{\text{pord}}} \left[x_9\right]^{P_\text{max}}_{P_\text{min}} \\
\dot{x}_{10} &=
\begin{cases}
\left[
\left[
\dfrac{[x_9]^{P_\text{max}}_{P_\text{min}}}{[x_1]^{\infty}_{0.01} \ T_{\text{g}}}
\right]^{I_{\text{dmax}}}_{I_{\text{dmin}}}
- \dfrac{x_{10}}{T_{\text{g}}}
\right]^{\mathsf{rrpwr}}_{-\mathsf{rrpwr}} \text{if}\;\kappa_{3} = 0 \\[16pt]
\left[
\left[
\dfrac{x_5 \ [x_9]^{P_\text{max}}_{P_\text{min}}}{[x_1]^{\infty}_{0.01} \ T_{\text{g}}}
\right]^{I_{\text{dmax}}}_{I_{\text{dmin}}}
- \dfrac{x_{10}}{T_{\text{g}}}
\right]^{\mathsf{rrpwr}}_{-\mathsf{rrpwr}} \text{otherwise}
\end{cases} \label{eq:x10_dynamics} \\
I_{\text{qv}} &= k_{\text{qv}} \ \text{db}^{\text{dbd2}}_{\text{dbd1}}(V_{\text{ref0}} - x_1) \label{eq:iqv} \\
\Delta f' &= \text{db}^{\text{fbd1}}_{\text{fbd2}}(f_{\text{ref}} - x_6) \label{eq:deltaf} \\
\Delta P_{\text{droop}} &= \text{D}_{\text{up}} \ [\Delta f']^{0}_{-\infty} + \text{D}_{\text{up}} \ [\Delta f']^{\infty}_{0} \label{eq:dpdroop} \\
P_{\text{lim}} &= \left[ P_{\text{ref}} + \Delta P_{\text{droop}} - x_2 \right]^{f_\text{emax}}_{f_\text{emin}} \label{eq:plim} \\
P_{\text{a}} &= \left[ x_7 + k_{\text{pg}} \ P_{\text{lim}} \right]^{P_\text{max}}_{P_\text{min}} \label{eq:pa} \\
\bar{V} &= (E_{\text{d}} + jE_{\text{q}}) e^{j\theta} \\
E_{\text{q}} &= V_{\text{q}} + I_{\text{d}}X_{\text{e}} \\
E_{\text{d}} &= V_{\text{d}} - I_{\text{q}}X_{\text{e}} \\
P &= V_{\text{d}} I_{\text{d}} + V_{\text{q}} I_{\text{q}} \label{eq:P} \\
Q &= V_{\text{d}} I_{\text{q}} - V_{\text{q}} I_{\text{d}}. \label{eq:Q}
\end{align}


{\color{black}
One challenge in parameterizing the DER\_A model is that it contains highly nonlinear elements, including deadband and saturation functions.} These are often implemented via if–else logic or other piecewise constructs, as in \eqref{eq:x1_dynamics}--\eqref{eq:x10_dynamics}. Such discontinuities pose a significant problem for gradient-based estimation algorithms that rely on continuous sensitivities between parameters, states, and outputs. Crucially, ignoring or oversimplifying these nonlinear elements can compromise the {\color{black}DER\_A} model's integrity and lead to an inaccurate or misleading representation of control behavior, especially under grid disturbances \cite{8626538}. 


An effective way to address these limitations is to replace the discontinuous nonlinearities with smooth approximations. Such functions provide a single, continuous and differentiable formulation of the system dynamics, removing abrupt transitions while preserving the key nonlinear behavior inherent in the original model. Accordingly, we leverage the method proposed in \cite{VasquezPlaza2023} to replace the piecewise-defined elements in \eqref{eq:saturation}--\eqref{eq:deadband} with smooth functional approximations. Specifically, the smooth saturation function (SSF) is given by:
\begin{align}
\text{SSF}_{\ell}^{\mathsf{h}}(x) &= \lambda + \mu \left(
\frac{x- \lambda}{\mu} \
\left(1 + \left(\frac{x-\lambda}{\mu}\right)^k\right)^{-\frac{1}{k}}
\right), \label{eq-SSF}
\end{align}
where $\ell$ and $\mathsf{h}$ denote, respectively, the lower and upper limits, $\lambda = \frac{\mathsf{h}+\ell}{2}$ and $\mu =\frac{\mathsf{h}-\ell}{2}$. The smooth deadband function (SDBF) is given by:
\begin{align}
\text{SDBF}_{\ell}^{\mathsf{h}}(x) &= x - \lambda - \mu \left(
\frac{x- \lambda}{\mu} \
\left(1 + \left(\frac{x-\lambda}{\mu}\right)^k\right)^{-\frac{1}{k}}
\right). \label{eq.SDBF}
\end{align}

\noindent
We refer the interested reader to \cite{VasquezPlaza2023} for details on \eqref{eq-SSF}--\eqref{eq.SDBF} and subsequent application in the state space equations of the DER\_A model. We therefore replace all saturation and deadband operators in \eqref{eq:x1_dynamics}--\eqref{eq:pa} with the smoothing functions \eqref{eq-SSF}--\eqref{eq.SDBF}. However, it is essential to carefully consider the effect of the smooth saturation function on the Jacobian before proceeding with observability analysis and parameter estimation. Section \ref{subsec:observability_analysis} offers a more detailed discussion of this issue and proposes a solution.

To illustrate how different flag configurations influence the ability to estimate parameters, we define two operating cases. These cases represent control modes found in grid-connected DERs and are used consistently throughout this paper. \emph{Case~1} corresponds to DERs configured for voltage support while maintaining constant active and reactive power. This case captures the essential dynamics of voltage, active, and reactive power without enabling active power or frequency control functions. \emph{Case~2} reflects DERs operating with volt--var and frequency--watt control schemes. In this configuration, voltage, and real and reactive power control dynamics are all active, representing the behavior of advanced inverter-based resources under full dynamic control. An overview of these model configurations, including the necessary active states and associated parameters for each case, is presented in Table~\ref{tab:model_config}.
\begin{table}[!ht]
\centering
\caption{DER\_A model operating cases and configuration}
\label{tab:model_config}
\renewcommand{\arraystretch}{1.15}
\begin{tabular}{@{}p{1.73cm}p{3.16cm}p{3.02cm}@{}}
\toprule
& \textbf{Case 1} & \textbf{Case 2} \\
\midrule
$\kappa_{1}$, $\kappa_{2}$, $\kappa_{3}$, $\kappa_{4}$ & 0, 0, 0, 0 & 1, 1, 0, 0 \\
\midrule
Enabled controls & Voltage support only; active and reactive power & Voltage, frequency, active power, and reactive power \\
\midrule
Active states & 
$x_1$, $x_3$, $x_4$, $x_8$, $x_9$, $x_{10}$ & 
All except $x_5$ \\
\midrule
Parameters & 
$P_{\text{max}}$, $P_{\text{min}}$, $I_{\text{dmax}}$, $I_{\text{dmin}}$, $I_{\text{q$\mathsf{h}$}}$, $I_{\text{q$\ell$}}$, $k_\text{qv}$, $T_{\text{iq}}$, $T_{\text{pord}}$, $T_{\text{rv}}$, $T_{\text{g}}$, $I_{\text{qmax}}$, $I_{\text{qmin}}$, $\text{dbd1}$, $\text{dbd2}$, $dP_\text{max}$, $dP_\text{min}$, &
All Case 1 parameters plus $\text{fbd1}$, $\text{fbd2}$, $k_{\text{ig}}$, $k_{\text{pg}}$, $\text{D}_{\text{dn}}$, $\text{D}_{\text{up}}$, $T_{\text{p}}$, $T_{\text{rf}}$, $f_\text{emax}$, $f_\text{emin}$ \\
\midrule
Inputs & $V$, $V_{\text{ref}}$, $P_{\text{ref}}$, $Q_{\text{ref}}$ & $V$, $V_{\text{ref}}$, $\text{freq}$, $P_{\text{ref}}$, $f_{\text{ref}}$, $\text{pfaref}$ \\
\bottomrule
\end{tabular}
\end{table}

\section{Observability and parameter estimation} \label{sec.III}
We investigate one's ability to estimate parameters in the DER\_A model under the two representative operating conditions described in Table \ref{tab:model_config}. This is achieved by augmenting the system state vector to include the parameters. The observability \cite{Hermann1977} of the augmented state vector thus serves as a structural indicator of whether a parameter has a discernible influence on the output through the system dynamics.

A conceptual parallel may be drawn between \emph{observability} and \emph{identifiability}. Observability is a structural property that assesses whether the state (or, in our case, the augmented state) can be uniquely determined from output trajectories (or measurements). Identifiability, by contrast, concerns determining model parameters based on the system's input--output behavior. Identifiability proceeds without embedding parameters into the state vector; instead, it evaluates the rank of the output sensitivity matrix or its generalizations. In this sense, observability can be interpreted as a special case of local identifiability, where the initial conditions act as the unknown parameters to be estimated \cite{10852366}. Observability is, therefore, a necessary but not sufficient condition for identifiability.

For nonlinear systems such as the DER\_A {\color{black}model}, characterized by saturation, deadband operators, and flags, observability can vary significantly with the operating mode. Specifically, the activation of different control settings alters the subset of dynamic states that evolve over time, which invariably influence which parameters can affect the system outputs and be estimated.

Observability analysis of nonlinear dynamical systems has often been approached using tools from differential geometry, particularly Lie derivatives~\cite{Zhao2019, Ali}. In this study, we adopt the Lie derivative-based approach. A nonlinear system is said to be observable at a state \(\bm{x}_0\) if the observability matrix constructed from successive Lie derivatives of the output function evaluated at \(\bm{x} = \bm{x}_0\) has full rank \cite{Ali}. In other words, the system is either found observable or unobservable based on the rank of the observability matrix.

Consider a nonlinear dynamical system
\begin{align}
\dot{\bm{x}} &= \bm{f}(\bm{x}, \bm{u})
\label{eq:nonlinear_state} \\
\bm{y} &= \bm{h}(\bm{x}),
\label{eq:nonlinear_output}
\end{align}
where \( \bm{x} \in \mathbb{R}^n \) is the state vector, \( \bm{u} \in \mathbb{R}^m \) is the input vector, and \( \bm{y} \in \mathbb{R}^p \) is the measured output. The Lie derivatives of an output function \( \bm{h}(\bm{x}) \) with respect to the vector field \( \bm{f}(\bm{x}, \bm{u}) \) are defined recursively, with the zeroth-order being the output itself, that is, $L_{\bm{f}}^0 \bm{h} = \bm{h}(\bm{x})$. Higher-order Lie derivatives are computed recursively, that is,

\begin{align}
L_{\bm{f}}^k \bm{h} &= \frac{\partial \left(L_{\bm{f}}^{k-1} \bm{h} \right)}{\partial \bm{x}} \cdot \bm{f}(\bm{x}) 
= \sum_{i=1}^{n} \frac{\partial \left(L_{\bm{f}}^{k-1} \bm{h} \right)}{\partial x_i} f_i(\bm{x}).
\end{align}

\noindent
The observability matrix is obtained by taking the Jacobian of
\begin{align}
\mathcal{L}(\bm{x}) &=
\big[\, h_1(\bm{x}), \dots, h_p(\bm{x}), \,
L_{\bm f} h_1(\bm{x}), \dots, L_{\bm f}^{\,n-1} h_p(\bm{x}) \,\big]^{\!\top},
\end{align}
with respect to \( \bm{x} \), that is,
\begin{align}
\mathcal{O}(\bm{x}) = \frac{\partial \mathcal{L}(\bm{x})}{\partial \bm{x}}.
\end{align}
We implement this method under the two operating cases.

\subsubsection{Observability analysis under voltage support mode only}
We augment the states in Case 1 with their associated parameters (see Table \ref{tab:model_config}) to form an {\color{black}augmented} state vector $\bm{x}^{\prime}~\!\!\in~\!\mathbb{R}^{23}$. Using two measurement sets $\{V,\, P,\, Q\}$ and $\{V,\, I_{\text{d}},\, I_{\text{q}}\}$, we compute the observability matrix and observed it to be rank-deficient, i.e., $\operatorname{rank}(\mathcal{O})\!<\!23$. This indicates that the augmented state $\bm{x}^{\prime}$ is not observable. 

Note that {\color{black}the} larger measurement set $\{V,\, I_{\text{d}},\, I_{\text{q}},\, P,\, Q\}$ also yields a rank-deficient observability matrix. {\color{black}This is because the DER\_A model uses a $dq$-reference frame aligned with the terminal voltage, in which the direct-axis voltage component (\(V_{d}\)) is set to the voltage magnitude ($V$) and the quadrature-axis component \(V_{q}=0\). Hence, $P \approx V I_{\text{d}}$ and $Q \approx V I_{\text{q}}$, and the larger measurement set $\{V,\, I_{\text{d}},\, I_{\text{q}},\, P,\, Q\}$ does not contribute independent information.} We therefore focus on two practically distinct measurement sets: $\{V,\, P,\, Q\}$ and $\{V,\, I_{\text{d}},\, I_{\text{q}}\}$; this choice balances observability considerations with realistic measurement availability.

Since $\bm{x}^{\prime}$ is not observable, the question is whether it can be made observable by reducing the number of parameters used to augment $\bm{x}$. {\color{black}Table~\ref{tab:EstimatedParams} summarizes the parameters considered in prior works and their estimability under the cases examined in this paper.}
\begin{table}[!hb]
\centering
\caption{DER\_A model parameters considered in previous works}
\label{tab:EstimatedParams}
\renewcommand{\arraystretch}{1.2}
\setlength{\tabcolsep}{2.5pt}
\begin{tabular}{l c c c c c c c c c c c c c c c c}
\toprule
\textbf{Reference} 
& \rotatebox{90}{$T_{\text{rv}}$} 
& \rotatebox{90}{$T_{\text{pord}}$} 
& \rotatebox{90}{$T_{\text{p}}$} 
& \rotatebox{90}{$T_{\text{rf}}$} 
& \rotatebox{90}{$T_{\text{g}}$} 
& \rotatebox{90}{$T_{\text{iq}}$} 
& \rotatebox{90}{$T_{\text{v}}$}
& \rotatebox{90}{$\text{dbd1}$} 
& \rotatebox{90}{$\text{dbd2}$} 
& \rotatebox{90}{$k_{\text{qv}}$} 
& \rotatebox{90}{$k_{\text{ig}}$} 
& \rotatebox{90}{$k_{\text{pg}}$} 
& \rotatebox{90}{$\text{D}_{\text{dn}}$} 
& \rotatebox{90}{$\text{D}_{\text{up}}$} 
& \rotatebox{90}{$\text{fbd1}$} 
& \rotatebox{90}{$\text{fbd2}$} \\
\midrule
\cite{Foroutan2023} & $\sbullet$ & & & & & & & $\sbullet$ & $\sbullet$ & $\sbullet$ & & & \\
\cite{9637939} & $\sbullet$ & $\sbullet$ & $\sbullet$ & $\sbullet$ & $\sbullet$ &$\sbullet$ &$\sbullet$ & & & $\sbullet$ & $\sbullet$ & $\sbullet$ & & & & \\
\cite{MHE_Tonkoski} & $\sbullet$ & $\sbullet$ & $\sbullet$ & $\sbullet$ & $\sbullet$ & $\sbullet$ & & $\sbullet$ & $\sbullet$ & $\sbullet$ & $\sbullet$ & $\sbullet$ & $\sbullet$ &$\sbullet$ &$\sbullet$ &$\sbullet$ \\
Estimable?$^{\dagger}$ & \ding{51} & \ding{51} & \ding{51} & \ding{51} & \ding{51} & \ding{51} 
& \ding{91} & {\color{black}\ding{55}} & {\color{black}\ding{55}} & \ding{51} & \ding{51} & \ding{51} &\ding{51} &\ding{51} & {\color{black}\ding{55}} & {\color{black}\ding{55}} \\
\bottomrule \\
\multicolumn{17}{l}{$^{\dagger}$Based on the cases considered in this paper: (\ding{51}) Yes. ({\color{black}\ding{55}}) No. (\ding{91}) Not} \\
\multicolumn{17}{l}{\;\,determined in this paper, since Vtripflag $=0$ for the examined cases.} \\
\end{tabular}
\end{table}
A systematic method to evaluate how to decrease $\bm{x}^{\prime}$ involves applying a singular value decomposition to the matrix $\mathcal{O}(\bm{x}^{\prime})$ to identify the parameters associated with the weakest observable direction, as discussed next.

\begin{figure}[!t]
\centering
\includegraphics[width=.49\linewidth]{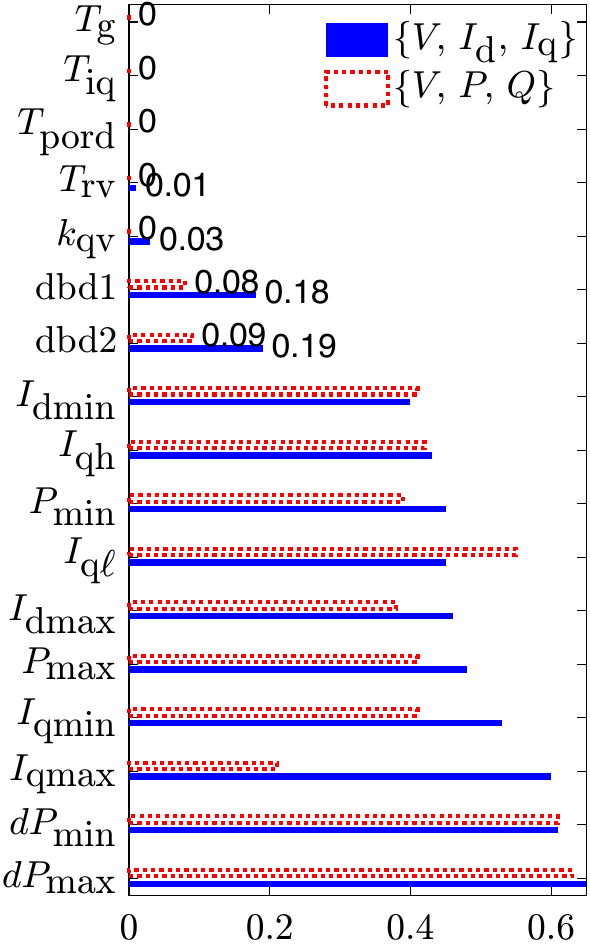}
\includegraphics[width=.49\linewidth]{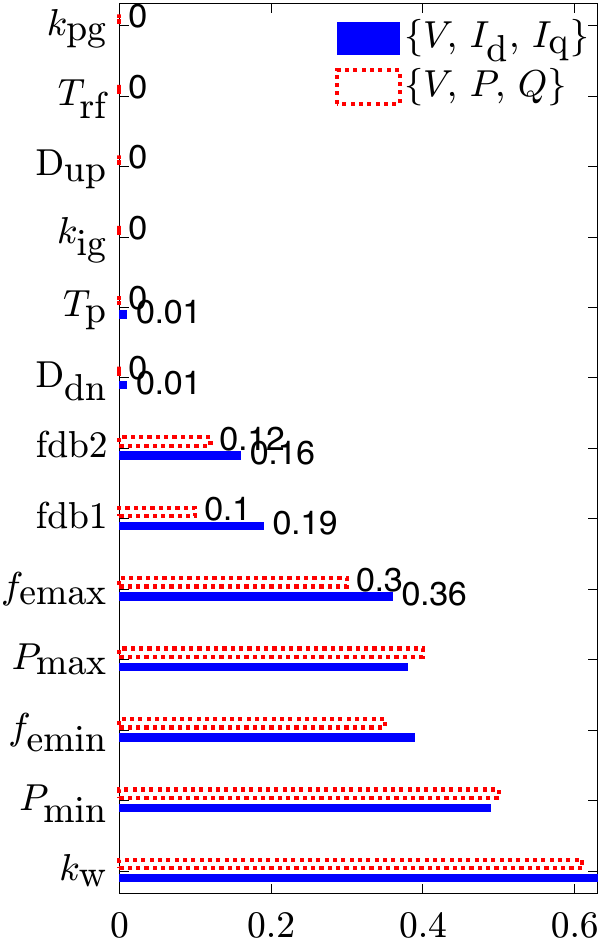}
\caption{Parameter weights in the least observable direction for Case 1 (left) and Case 2 (right).}
\label{fig:parameter weights}
\end{figure}

The analysis focuses on the smallest nonzero singular value, whose corresponding right singular vector defines the direction in the augmented state space where observability is weakest \cite{MADTHARAD200399}. We normalize the absolute values of the parameter components in {\color{black}the right singular} vector to illustrate the contribution of each parameter to that direction. Parameters with larger values are strongly aligned with the weakest observable subspace, indicating limited influence on the measured outputs; see Fig.~\ref{fig:parameter weights}.

{\color{black}
We emphasize that the primary criterion is the rank of the observability matrix of the augmented system. When the matrix is rank-deficient, the augmented state is not locally observable, indicating a structural limitation in uniquely determining it from the available measurements. The singular value decomposition is used to analyze the directions associated with this rank deficiency and to guide the reduction of the augmented parameter set. In this context, the smallest singular value serves as a measure of local observability strength and conditioning along the trajectory, rather than a standalone indicator of individual parameter identifiability. Consequently, a parameter's alignment with a weakly observable direction may reflect correlation with other states or parameters, rather than a complete lack of identifiability.}
From Fig.~\ref{fig:parameter weights}, the parameters associated with saturation blocks are the least observable. However, we note that inverter manufacturers determine saturation via firmware, and those values are fixed \cite{NERC2026, MHE_Tonkoski}. Moreover, the saturation values for the DER\_A {\color{black}model} are generally set according to industry standards. The maximum current limit is set at $I_{\text{max}} = 1.2$~pu, which corresponds to the 110--120\% overcurrent headroom of modern inverter-based DERs. The reactive current limit is set as $I_{\text{q}\mathsf{h}1}/I_{\text{q}\ell{1}} = \pm 1.0$ pu. Also, given that the ramp-rate \( x_8 \) is largely determined by the weakly observable saturation parameters $dP_\text{max}$ and $dP_\text{min}$, and is not a direct measurable physical quantity, it is excluded from the state vector. By removing saturation bounds and other threshold-type parameters, the augmented state vector is reduced to the set as given in \eqref{eq:reduced state vector}, and this results in a full-rank \(\mathcal{O}\) matrix.
\begin{align}
{\color{black}\bm{x}^{\prime}_{\text{case 1}}} = [
& x_1,\ x_3,\ x_4,\ x_9,\ x_{10},\ T_{\text{rv}},\ k_{\text{qv}},\ T_{\text{g}},\ T_{\text{iq}},\ T_{\text{pord}}]^\top.
\label{eq:reduced state vector}
\end{align}
\subsubsection{Observability analysis under reactive power--voltage control and active power--frequency control} 
Similarly, we remove the associated breakpoints and deadbands from the set of parameters associated with the controls described in Case 2. The resulting augmented state is given by:
\begin{align}
{\color{black}\bm{x}^{\prime}_{\text{case 2}}} = \big[\, 
& x_1,\ x_2,\ x_3,\ x_4,\ x_6,\ x_7,\ x_9,\ x_{10},\ T_{\text{p}},\ T_{\text{rf}}, \notag \\
& k_{\text{pg}},\ k_{\text{ig}},\ \text{D}_{\text{dn}},\ \text{D}_{\text{up}}\, \big]^\top. 
\label{eq:reduced_state_vector2}
\end{align}

\begin{figure}[!t]
\centering
\includegraphics[width=0.8\linewidth]{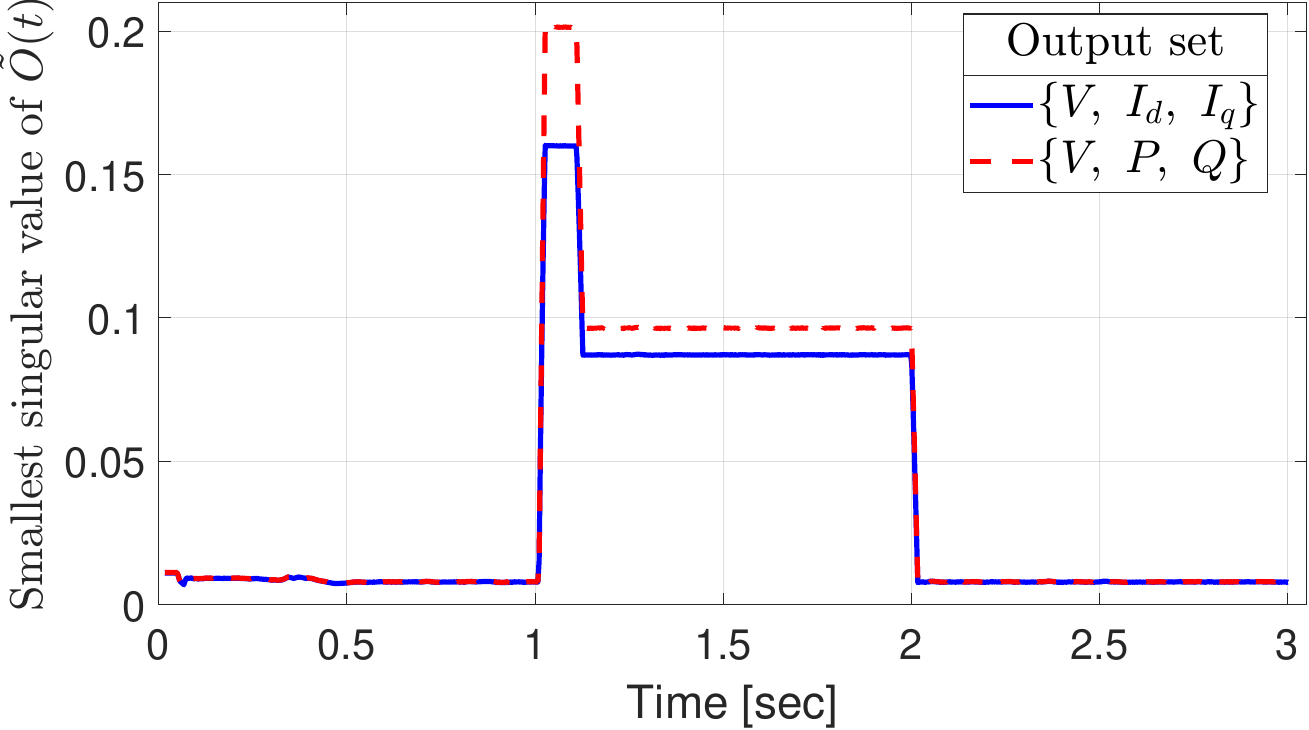}
\caption{Smallest singular value of the observability matrix.}
\label{fig:smallest singular observability matrix}
\end{figure}

\noindent
{\color{black}To evaluate the observability strength across the different measurement configurations, we assess local observability along the simulated system trajectory by computing, at each time instant, the smallest singular value of the observability matrix; see Fig.~\ref{fig:smallest singular observability matrix}. Because we consider local observability, this yields a time-varying measure of observability strength. The mean and standard deviation of this quantity are then used as summary metrics.} From Table~\ref{tab:observability_summary}, the measurement set $\{V,\, P,\, Q\}$ yields a higher mean singular value compared to $\{V,\, I_{\text{d}},\, I_{\text{q}}\}$, which indicates a stronger level of observability for the system dynamics. This suggests that $\{V,\, P,\, Q\}$ is a better choice for the parameter estimation problem in this case.

\begin{remark}
This paper does not seek to provide a definitive answer on the optimal measurement set. The answer lies with the system operator, as different distribution networks will (i) have access to disparate measurement sets and (ii) yield specific flag configurations in the DER\_A model based on the DERs in the field. Instead, our goal is to offer researchers and practitioners a principled approach for addressing the DER\_A {\color{black}model} parameterization. The MATLAB code to replicate all results in this paper is available \cite{code}.
\end{remark}

\begin{remark}
{\color{black}
The estimable parameter subsets identified in this work are associated with the operating conditions, disturbance profile, and measurement sets considered. Local observability in nonlinear systems depends on the operating point and level of excitation; consequently, parameter rankings may vary under different system conditions, including alternative disturbances, feeder configurations, or DER penetration levels.} 
\end{remark}

\begin{table}[!t]
\centering
\caption{The mean and the standard deviation of the smallest singular value of the observability matrix}
\begin{tabular}{ccc}
\toprule
\textbf{Measurement set} & \textbf{Mean} & \textbf{Standard deviation} \\
\midrule
$\{V,\, I_{\text{d}},\, I_{\text{q}}\}$ & 0.0369 & 0.0423 \\
$\{V,\, P,\, Q\}$ & 0.0410 & 0.0495 \\
\bottomrule
\end{tabular}
\label{tab:observability_summary}
\end{table}

It is important to note that we do not estimate control flags as they are not continuous-valued tunable variables. In view of these considerations, the parameters selected for estimation in this work are those in \eqref{eq:reduced state vector}--\eqref{eq:reduced_state_vector2}.

\subsection{Impact of the SSF derivative on the Jacobian numerical stability and observability}\label{subsec:observability_analysis}
\begin{figure}[!t]
\centering
\includegraphics[width=0.9\linewidth]{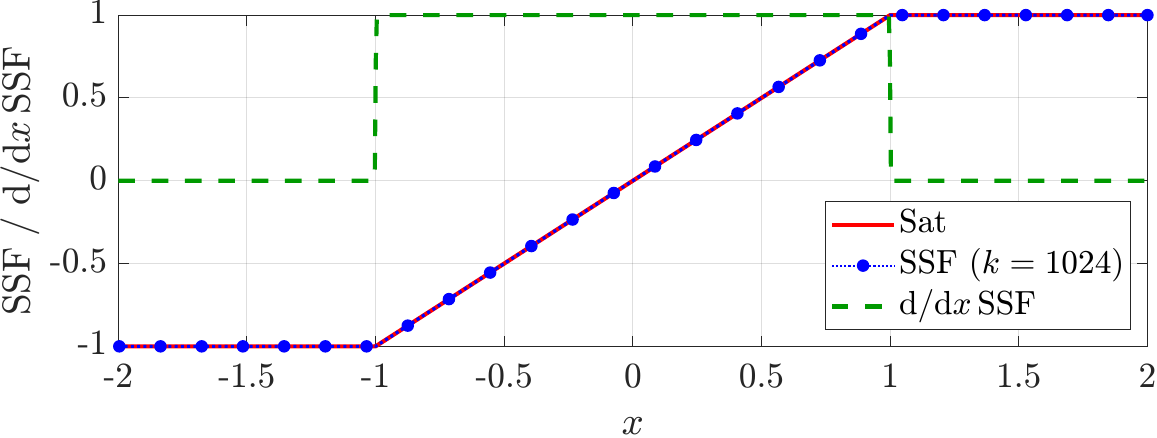}
\caption{Smooth saturation function and its derivative.}
\label{fig:SSF}
\end{figure}

\begin{figure}[!t]
\centering
\includegraphics[width=0.9\linewidth]{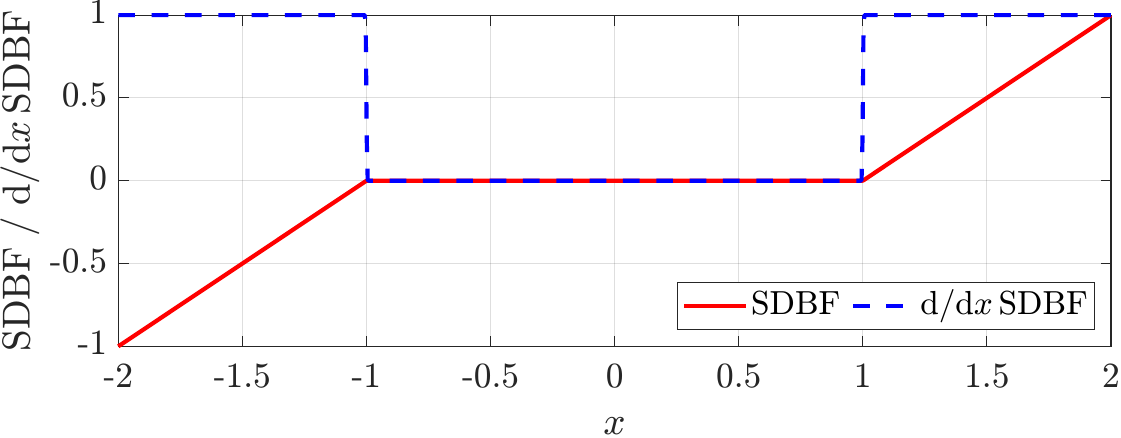}
\caption{Smooth deadband function and its derivative.}
\label{fig:SDBF}
\end{figure}

The work in \cite{MHE_Tonkoski} reported that only three parameters could be successfully estimated using the EKF. According to the authors \cite{MHE_Tonkoski}, attempts to estimate additional parameters resulted in divergence, primarily due to the nonlinearities introduced by deadbands and saturation functions in the {\color{black}DER\_A} model. Here, we discuss the mathematical structure of the SSF derivative, its effect on system observability and Jacobian numerical stability, and our approach to addressing these numerical issues. The derivative of \eqref{eq-SSF} is given by
\begin{equation}
\frac{d}{dx} \text{SSF}_{\ell}^{\mathsf{h}}(x) = \left(1 + z^k \right)^{-1 - \frac{1}{k}},
\end{equation}
with \( z = \frac{x - \lambda}{\mu} \). From Fig.~\ref{fig:SSF}, the derivative reaches its peak value of approximately 1 at \( x = \lambda \), where the function transitions smoothly between its linear and saturated regions. However, for inputs \( x \) such that \( |x - \lambda| \gg \mu \), the argument \( z^k \) becomes very large, and this causes the derivative to approach zero:
\begin{equation}
\lim_{|z| \to \infty} \frac{d}{dx} \text{SSF}_{\ell}^{\mathsf{h}}(x) \rightarrow 0.
\end{equation}
By extension, this behavior also applies to the smooth deadband function shown in Fig.~\ref{fig:SDBF}. This behavior closely resembles the vanishing gradient problem observed in the sigmoid and hyperbolic tangent functions widely used in neural networks \cite{Vanishinggradient}. Consequently, the SSF exhibit negligible gradient magnitudes in their saturated tails, leading to diminished sensitivity to input variations. In the context of nonlinear estimation methods, including Kalman filtering, this presents a fundamental challenge.
For instance, the EKF relies on a Jacobian matrix such that when \( \frac{d}{dx} \text{SSF}_{\ell}^{\mathsf{h}}(x) \approx 0 \), the Jacobian becomes nearly rank-deficient, which suppresses state observability and compromises the innovation vector in the filtering step. As the Jacobian entries approach zero, the Kalman gain leads to negligible corrections to the predicted state and potential filter divergence. Furthermore, this may cause an ill-conditioned Jacobian in the EKF, or non-positive definite covariance matrices during Cholesky decompositions in other filters such as the unscented Kalman filter (UKF). To mitigate this, a common strategy involves enforcing a lower bound on the derivative magnitude. Formally, 
\begin{equation}
\frac{d}{dx} \text{SSF}_{\ell}^{\mathsf{h}}(x) \gets \max\left( \epsilon, \left(1 + z^k \right)^{-1 - \frac{1}{k}} \right),
\end{equation}
where \( \epsilon \) is a small threshold such as \( \epsilon = 10^{-6} \). This preserves numerical stability by preventing the Jacobian entries (i.e., the partial derivatives of the state equations) from becoming zero, while retaining the overall shape and saturating behavior of the SSF function. {\color{black}This modification can be viewed as a form of numerical regularization, which may introduce a minor trade-off between model fidelity in the saturated region of a curve and numerical stability during estimation.}

\subsection{DER\_A model parameter estimation}
To tune the parameters of the DER\_A model, we adopt a Kalman filter-based estimation framework using an augmented state formulation. The selected parameters are appended to the system’s state vector and jointly estimated over time using voltage and power measurements. It is essential to clarify that these parameters are often not entirely unknown, but rather initialized from nominal values and refined based on their fit to system response data. If no previous information is available, the values suggested in \cite{NERC2026} can be used for initialization.

We define the \textit{augmented state vector} as follows:
\begin{align}
\bm{x}_k' = \big[\, \bm{x}_k \;\; \bm{\theta}_{k} \,\big]^\top \in \mathbb{R}^{n+p},
\end{align}

\noindent
where $\bm{x}_k \in \mathbb{R}^n$ is the state vector at discrete time $k\in\mathbb{Z}_{>0}$, and $\bm{\theta}_{k} \in \mathbb{R}^p$ is the vector of uncalibrated parameters. The discrete-time system dynamics can be expressed as
\begin{align}
\bm{x}^{\prime}_{k+1} &= \bm{f}(\bm{x}^{\prime}_k, \bm{u}_k) + \bm{w}_k \\
\bm{z}_k &= \bm{h}(\bm{x}^{\prime}_k) + \bm{v}_k,
\end{align}

\noindent
where $\bm{f}(\cdot)$ is the vector-valued nonlinear process model that maps $\bm{x}^{\prime}_k$ to $\bm{x}^{\prime}_{k+1}${\color{black}, with parameters modeled as $\bm{\theta}_{k+1} = \bm{\theta}_k$;} $\bm{h}(\cdot)$ is a vector-valued nonlinear measurement function; and $\bm{w}_k \sim \mathcal{N}(0, \bm{W}_k)$ and $\bm{v}_k \sim \mathcal{N}(0, \bm{R}_k)$ are the process and measurement noise, respectively. We assume that $\bm{w}_k$ and $\bm{v}_k$ are independent, zero-mean, white Gaussian noise processes, i.e.,
$\mathbb{E}[\bm{w}_k] = 0$, $\mathbb{E}[\bm{w}_k \bm{w}_k^\top] = \bm{W}_k$, 
$\mathbb{E}[\bm{v}_k] = 0$, $\mathbb{E}[\bm{v}_k \bm{v}_k^\top] = \bm{R}_k$, 
$\mathbb{E}[\bm{v}_j \bm{v}_k^\top] = \mathbb{E}[\bm{w}_k \bm{w}_j^\top] = \bm{0}$, $j \ne k$, 
and $\mathbb{E}[\bm{w}_k \bm{v}_j^\top] = \bm{0}$.

{\color{black}
For numerical implementation, $\bm{f}(\cdot)$ is obtained by discretizing the continuous-time DER\_A dynamics using a fourth-order Runge--Kutta scheme with sampling interval $\Delta t$ = 0.02s. The augmented state is propagated as
\begin{equation}
\boldsymbol{x}'_{k+1}
=
\boldsymbol{x}'_k
+\frac{1}{6}\left(\boldsymbol{k}_1 + 2\boldsymbol{k}_2 + 2\boldsymbol{k}_3 + \boldsymbol{k}_4\right),
\end{equation}
where $\boldsymbol{k}_1,\boldsymbol{k}_2,\boldsymbol{k}_3,\boldsymbol{k}_4$ are the standard Runge--Kutta stage evaluations of the augmented system dynamics.

Following \cite{MarcosIEKF}, the EKF employs first-order linearization of $\bm{f}(\cdot)$ and $\bm{h}(\cdot)$ about the current estimate. The covariance prediction is given by
\begin{equation}
\boldsymbol{P}_{k|k-1}
=
\boldsymbol{F}_{k-1}\boldsymbol{P}_{k-1|k-1}\boldsymbol{F}_{k-1}^{\top}
+
\boldsymbol{Q}_k,
\end{equation}
where $\boldsymbol{F}_{k-1} = \left.\partial f(\boldsymbol{x}', \boldsymbol{u}_{k-1}) / \partial \boldsymbol{x}'\right|_{\boldsymbol{x}'=\hat{\boldsymbol{x}}'_{k-1|k-1}}$ is the Jacobian matrix. The correction step is computed as follows:
\begin{align}
\boldsymbol{K}_k
&=
\boldsymbol{P}_{k|k-1}\boldsymbol{H}_k^{\top}
\left(
\boldsymbol{H}_k\boldsymbol{P}_{k|k-1}\boldsymbol{H}_k^{\top}
+
\boldsymbol{R}_k
\right)^{-1} \\
\hat{\boldsymbol{x}}'_{k|k}
&=
\hat{\boldsymbol{x}}'_{k|k-1}
+
\boldsymbol{K}_k\left(\boldsymbol{z}_k - h(\hat{\boldsymbol{x}}'_{k|k-1})\right) \\
\boldsymbol{P}_{k|k}
&=
(\boldsymbol{I} - \boldsymbol{K}_k \boldsymbol{H}_k)\boldsymbol{P}_{k|k-1},
\end{align}
where $\boldsymbol{H}_k = \left.\partial h / \partial \boldsymbol{x}'\right|_{\boldsymbol{x}'=\hat{\boldsymbol{x}}'_{k|k-1}}.$

On the other hand, UKF uses the unscented transform to select a set of sigma points that explicitly capture the first- and second-order moments of the state distribution. These points are propagated through the nonlinear function to approximate the posterior mean and covariance \cite{Simon2006}. Let $n_a = n + p$ denote the dimension of the augmented state. A set of $2n_a + 1$ sigma points $\{ \chi_{k-1}^{(i)} \}_{i=0}^{2n_a}$ is generated from $(\hat{x}_{k-1|k-1}, P_{k-1|k-1})$ and propagated through $\bm{f}(\cdot)$ such that
\begin{equation}
\boldsymbol{\chi}^{(i)}_{k|k-1} = f(\boldsymbol{\chi}^{(i)}_{k-1}, \boldsymbol{u}_{k-1}).
\end{equation}

The predicted mean and covariance are computed as
\begin{align}
\hat{\boldsymbol{x}}'_{k|k-1}
&=
\sum_{i=0}^{2n_a} W_i^{(m)} \boldsymbol{\chi}^{(i)}_{k|k-1} \\
\boldsymbol{P}_{k|k-1}
&=
\sum_{i=0}^{2n_a}
W_i^{(c)}
\left(\boldsymbol{\chi}^{(i)}_{k|k-1} - \hat{\boldsymbol{x}}_{k|k-1}\right)
\left(\boldsymbol{\chi}^{(i)}_{k|k-1} - \hat{\boldsymbol{x}}_{k|k-1}\right)^{\top} \notag \\
&\quad + \boldsymbol{Q}_k.
\end{align}
where $W_i^{(m)}$ and $W_i^{(c)}$ denote the weights associated with the mean and covariance, respectively. The transformed sigma points in the measurement space are given by
\begin{equation}
\boldsymbol{z}^{(i)}_{k|k-1} = h(\boldsymbol{\chi}^{(i)}_{k|k-1}),
\end{equation}

\noindent with predicted measurement mean
\begin{equation}
\hat{\boldsymbol{z}}_{k|k-1}
=
\sum_{i=0}^{2n_a} W_i^{(m)} \boldsymbol{z}^{(i)}_{k|k-1},
\end{equation}

\noindent and covariance matrices
\begin{align}
\boldsymbol{P}_{zz,k}
&=
\sum_{i=0}^{2n_a}
W_i^{(c)}
\left(\boldsymbol{z}^{(i)}_{k|k-1} - \hat{\boldsymbol{z}}_{k|k-1}\right)
\left(\boldsymbol{z}^{(i)}_{k|k-1} - \hat{\boldsymbol{z}}_{k|k-1}\right)^{\top} \nonumber \\
&\quad + \boldsymbol{R}_k, \\
\boldsymbol{P}_{xz,k}
&=
\sum_{i=0}^{2n_a}
W_i^{(c)}
\left(\boldsymbol{\chi}^{(i)}_{k|k-1} - \hat{\boldsymbol{x}}_{k|k-1}\right)
\left(\boldsymbol{z}^{(i)}_{k|k-1} - \hat{\boldsymbol{z}}_{k|k-1}\right)^{\top}.
\end{align}

The Kalman gain is obtained using
\begin{equation}
\boldsymbol{K}_k = \boldsymbol{P}_{xz,k}\boldsymbol{P}_{zz,k}^{-1},
\end{equation}

with posterior update
\begin{align}
\hat{\boldsymbol{x}}'_{k|k}
&=
\hat{\boldsymbol{x}}'_{k|k-1}
+
\boldsymbol{K}_k\left(\boldsymbol{z}_k - \hat{\boldsymbol{z}}_{k|k-1}\right) \\
\boldsymbol{P}_{k|k}
&=
\boldsymbol{P}_{k|k-1}
-
\boldsymbol{K}_k\boldsymbol{P}_{zz,k}\boldsymbol{K}_k^{\top}.
\end{align}

In both implementations, the process and measurement noise covariance matrices $\boldsymbol{Q}_k$ and $\boldsymbol{R}_k$ are chosen as diagonal matrices, with reduced variance assigned to the parameter components of $\boldsymbol{Q}_k$ relative to the dynamic states. For the UKF, the sigma-point construction follows the standard unscented transform with parameters $\alpha = 10^{-3}$ and $\beta = 2$, which are widely adopted for Gaussian priors in nonlinear state estimation \cite{Simon2006}.
}

\section{Numerical experiment setup and results}\label{sec.IV}
We utilize the IEEE 34-node test feeder, modified to include solar PV generation with a total installed capacity of 500~kW. Five identical solar PV units are connected near the feeder head to ensure that the dynamic response of each inverter is clearly captured in the measurements at the point of common coupling. This placement minimizes the masking effects of line impedance and intermediary control devices \cite{BALFOUR2021111067}. The solar PV systems specifications are summarized in Table~\ref{tab:pv_specs}.
\begin{table}[!t]
\centering
\caption{Specifications of each solar PV system}
\label{tab:pv_specs}
\begin{tabular}{l l} \toprule
\textbf{Component } & \textbf{Value / Description} \\ \midrule
PV Array Model & 330 $\times$ SunPower SPR-305 modules \\
Rated Power & 100 kW \\
Irradiance & 1000 W/m$^2$ \\
Temperature Input & 25$^\circ$C \\
MPPT Control & Incremental conductance + integral regulator \\
Boost Converter & 5 kHz switching frequency, 500 V output \\
Inverter Type & 3-level voltage source converter) \\ \bottomrule
\end{tabular}
\end{table}

\begin{figure}[!t]
\centering
\includegraphics[width=0.9
\linewidth]{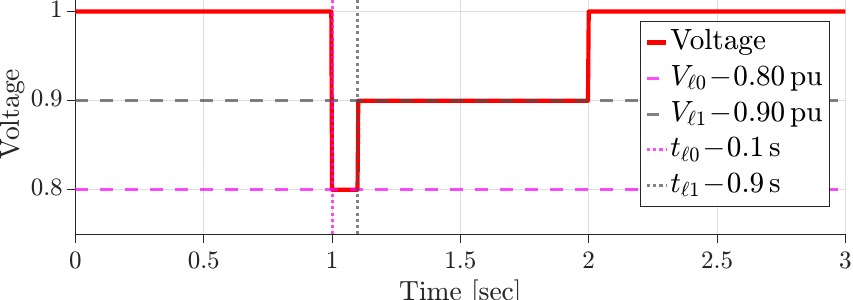}
\caption{Substation voltage profile.}
\label{fig:substation profile}
\end{figure}

\begin{figure}[ht]
\centering
\includegraphics[width=1\linewidth]{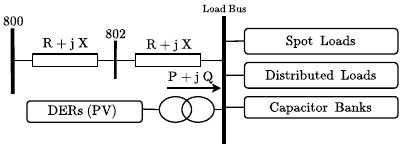}
\vspace{-1em}
\caption{Equivalent IEEE 34-node test feeder diagram.}
\label{testfeeder}
\end{figure}

Each solar PV system is modeled as a grid-connected unit composed of a PV array with frequency--watt and volt--var support. A maximum power point tracking (MPPT) controller extracts maximum energy from the irradiance input and regulates a boost converter, which raises the DC link voltage. The DC output is converted to AC through a three-level voltage source converter and interfaced with the utility grid via a step-up transformer. Irradiance and temperature are provided as inputs to determine the instantaneous solar PV output power. The MPPT controller determines $P_{\text{ref}}$, while $Q_{\text{ref}}$ is fixed to inject 0.2 pu of reactive power. Following \cite{EPRI2019}, we design the substation node 800 voltage profile using \eqref{eq-voltage eq} to induce dynamics; see Fig. \ref{fig:substation profile}.
\begin{align}
V(t) = 
\begin{cases}
a & 1 \leq t < 1 + \frac{b}{60} \\[6pt]
\displaystyle
\frac{d - a}{9} \left(t - 1 - \frac{b}{60}\right) + a & 
1 + \frac{b}{60} \leq t < 1 + c \\[6pt]
1.0 & \text{otherwise}.
\end{cases}
\label{eq-voltage eq}
\end{align}
where \( a = 0.80 \) is the sag value observed from 1.0 to 1.1 s, \( b = 60 \) represents 60 cycles at 60 Hz, equivalent to 0.10 s of sag duration, \( c = 0.90 \) is the ramp-hold duration corresponding to recovery from 1.1 to 2.0 s, and \( d = 0.90 \) indicates the ramp starting point, with voltage linearly increasing from 0.90 to 1.0 pu. The IEEE 34-node test feeder is thus modeled as a 2.5 MVA, 24.9 kV system with an aggregate 20\% solar PV penetration. An equivalent circuit is shown in Fig.~\ref{testfeeder}.

\begin{table}[!t]
\centering
\caption{Comparison of parameter estimates}
\label{tab:ukf_ekf_cov}
\begin{tabular}{lcccc}
\toprule
\multicolumn{1}{c}{Parameter} & \multicolumn{2}{c}{UKF} & \multicolumn{2}{c}{EKF}\\
\cmidrule(lr){2-3}\cmidrule(lr){4-5}
& Estimate & CoV & Estimate & CoV \\
\midrule
$T_{\text{rv}}$ & 0.190 & $5.5e-3$ & 0.210 & $5.3e-3$ \\
$k_{\text{qv}}$ & 4.950 & $1.0e-3$ & 4.940 & $1.3e-3$\\
$T_{\text{g}}$ & 0.210 & $3.6e-3$ & 0.198 & $3.2e-3$ \\
$T_{\text{iq}}$ & 0.019 & $4.3e-3$ & 0.020 & $4.2e-3$ \\
$T_{\text{pord}}$ & 0.019 & $4.0e-3$ & 0.020 & $4.1e-3$\\
$T_{\text{p}}$ & 0.020 & $3.9e-3$ & 0.020 & $3.9e-3$\\
$k_{\text{pg}}$ & 0.502 & $3.7e-6$ & 0.500 & $3.7e-6$ \\
$k_{\text{ig}}$ & 9.972 & $6.0e-7$ & 9.970 & $6.0e-7$ \\
$T_{\text{rf}}$ & 0.033 & $7.8e-3$ & 0.030 & $7.8e-3$\\
$D_{\text{dn}}$ & 20.08 & $7.4e-6$ & 20.10 & $7.6e-6$ \\
$D_{\text{up}}$ & 19.98 & $7.5e-6$ & 19.97 & $7.4e-3$\\
\bottomrule
\end{tabular}
\end{table}

{\color{black}\subsection{Estimation results under the baseline voltage profile}}
The EKF and UKF are executed over a 3-second simulation period, with measurements reported at a rate matching current North American practice for synchrophasor measurements: 30 samples per second. {\color{black}We remark that the measurements considered in this work are consistent with dynamic data \cite{NERC2023_data_collection} collected by transmission operators.}

The results in Table \ref{tab:ukf_ekf_cov} indicate that the EKF and UKF converged to similar estimated parameter values. The coefficient of variation (CoV) is calculated as the standard deviation divided by the mean estimate; lower values signify less relative uncertainty. {\color{black}Table \ref{tab:parameter_estimates} further compares the parameter values suggested by the NERC reliability guideline, the initial values used for the estimation, and the calibrated values obtained from the proposed framework.}

\begin{table}[!t]
\centering
\caption{Parameter calibration results}
\label{tab:parameter_estimates}
\begin{tabular}{lccc}
\toprule
 & \textbf{Value suggested} & \multicolumn{2}{c}{\textbf{Calibration}} \\ \cline{3-4}
\textbf{Parameter} & \textbf{by NERC guideline} & \textbf{Initialization} & \textbf{Estimated value} \\
\midrule
$T_{\text{rv}}$ & 0.02 & 0.01 & 0.2100\\
$k_{\text{qv}}$ & 5.00 & 4.00 & 4.950 \\
$T_{\text{g}}$ & 0.02 & 0.012 & 0.019 \\
$T_{\text{iq}}$ & 0.02 & 0.06 & 0.0201 \\
$T_{\text{pord}}$ & 0.02 & 0.01 & 0.0206 \\
$T_{\text{p}}$ & 0.02 & 0.01 & 0.0201 \\
$k_{\text{pg}}$ & 0.10 & 0.45 & 0.500 \\
$k_{\text{ig}}$ & 10.00 & 7.00 & 9.970 \\
$T_{\text{rf}}$ & 0.02 & 0.01 & 0.030 \\
$D_{\text{dn}}$ & 20.00 & 15.00 & 20.100 \\
$D_{\text{up}}$ & 20.00 & 18.00 & 19.980 \\
\bottomrule
\end{tabular}
\end{table}

\begin{figure}[!t]
\centering
\includegraphics[width=0.9\linewidth]{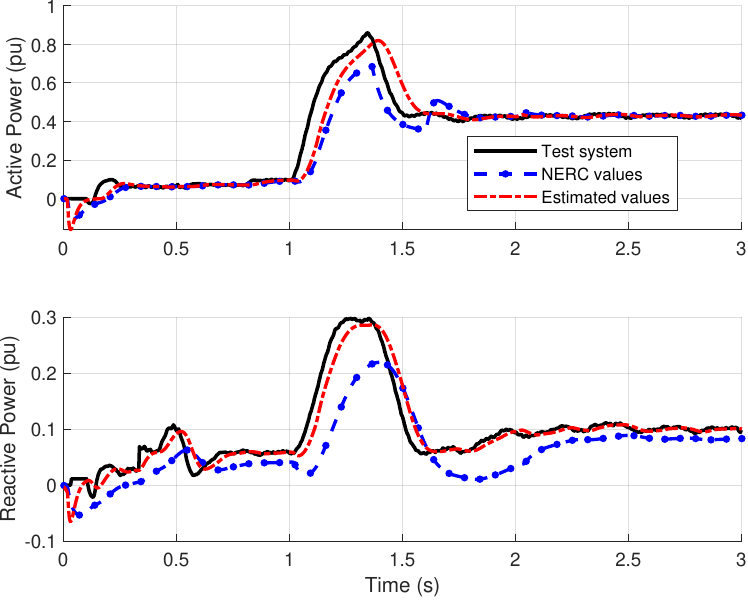}
\caption{Case1 - Comparison of simulated and measured active and reactive power using the DER\_A model and test system, respectively, with measurement set $\{V,\,P,\,Q\}$.}
\label{fig: power comparison}
\end{figure}

\begin{figure}[!t]
\centering
\includegraphics[width=0.9\linewidth]{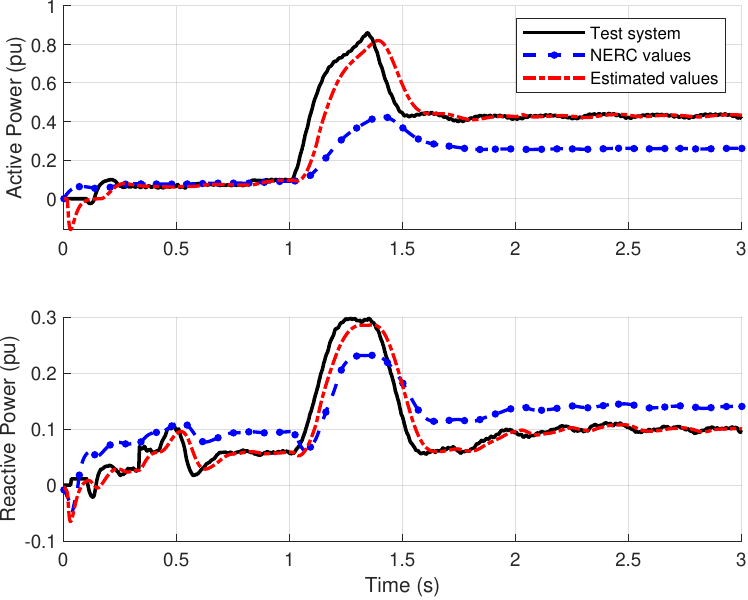}
\caption{Case 2 - Comparison of simulated and measured active and reactive power using the DER\_A model and test system, respectively, with measurement set $\{V,\,P,\,Q\}$.}
\label{fig: power comparison case 2}
\end{figure}

To verify the accuracy of the estimated parameters, Figs. \ref{fig: power comparison} and \ref{fig: power comparison case 2} compare the active and reactive power of the DER\_A model calibrated using the EKF with that of the test system. It also presents results for the case in which the parameter values from the NERC reliability guideline \cite{NERC2026} are used without model calibration. Similarly, the UKF was tested under the same conditions and showed convergence behavior comparable to that of the EKF.

\begin{figure*}[ht]
\centering
\setlength{\tabcolsep}{1pt}
\renewcommand{\arraystretch}{1}
\begin{tabular}{ccc}
\begin{subfigure}[b]{0.33\textwidth}\centering
\includegraphics[width=\textwidth]{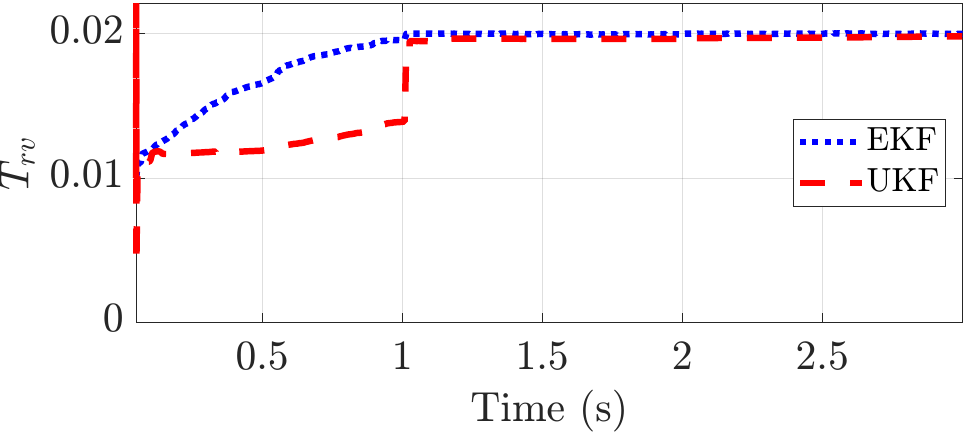}
\end{subfigure} &
\begin{subfigure}[b]{0.33\textwidth}\centering
\includegraphics[width=\textwidth]{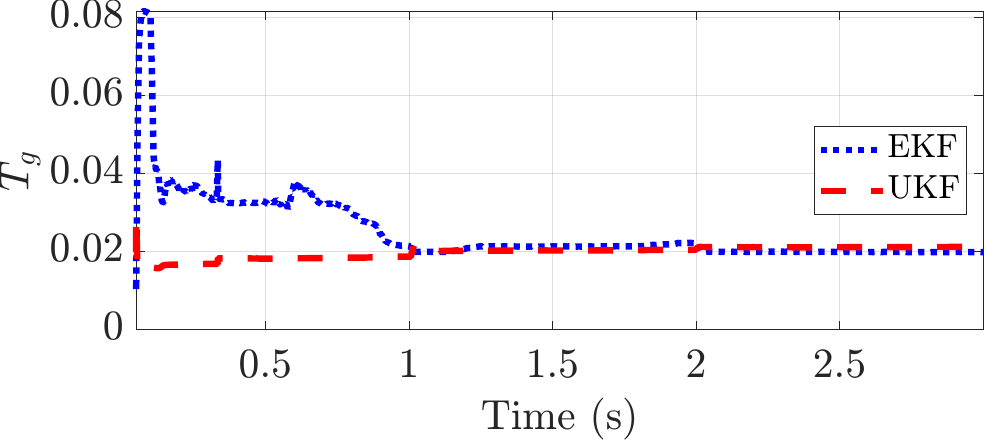}
\end{subfigure} &
\begin{subfigure}[b]{0.33\textwidth}\centering
\includegraphics[width=\textwidth]{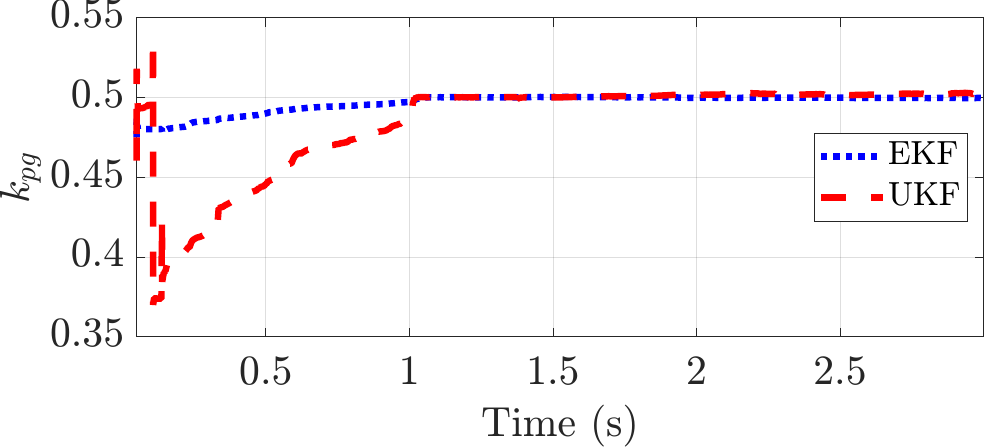}
\end{subfigure} \\[3pt]
\begin{subfigure}[b]{0.33\textwidth}\centering
\includegraphics[width=\textwidth]{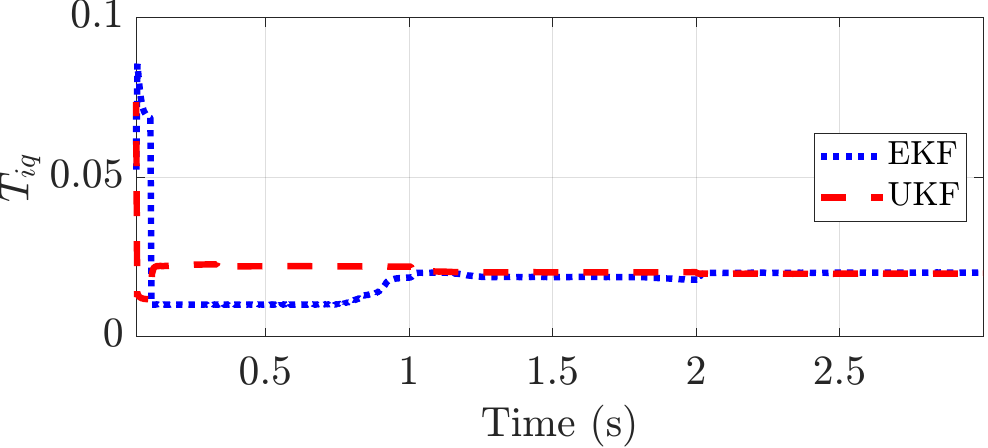}
\end{subfigure} &
\begin{subfigure}[b]{0.33\textwidth}\centering
\includegraphics[width=\textwidth]{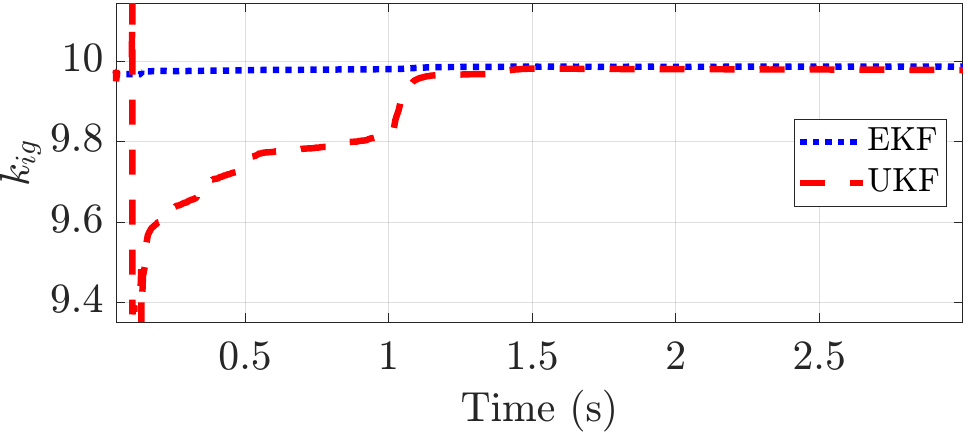}
\end{subfigure} &
\begin{subfigure}[b]{0.33\textwidth}\centering
\includegraphics[width=\textwidth]{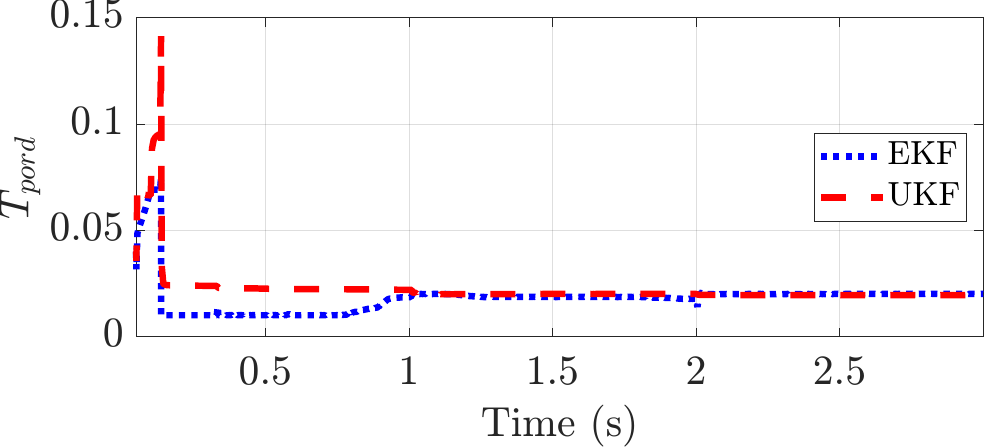}
\end{subfigure} \\
\end{tabular}
\caption{Convergence behavior of EKF/UKF parameter estimates.}
\label{fig:EKF_UKF_parameters}
\end{figure*}

This cross-validation confirms the effectiveness and consistency of Kalman filtering-based methods. Specifically, these findings support the use of EKF or UKF for automatic parameter calibration in aggregated DER models, especially when limited prior knowledge is available or when parameter drift occurs over time. The trajectories of six estimated parameters are shown in Fig. \ref{fig:EKF_UKF_parameters}, with additional plots accessible in the shared \texttt{MATLAB} code \cite{code}.

\begin{figure}[!t]
\centering
\includegraphics[width=1\linewidth]{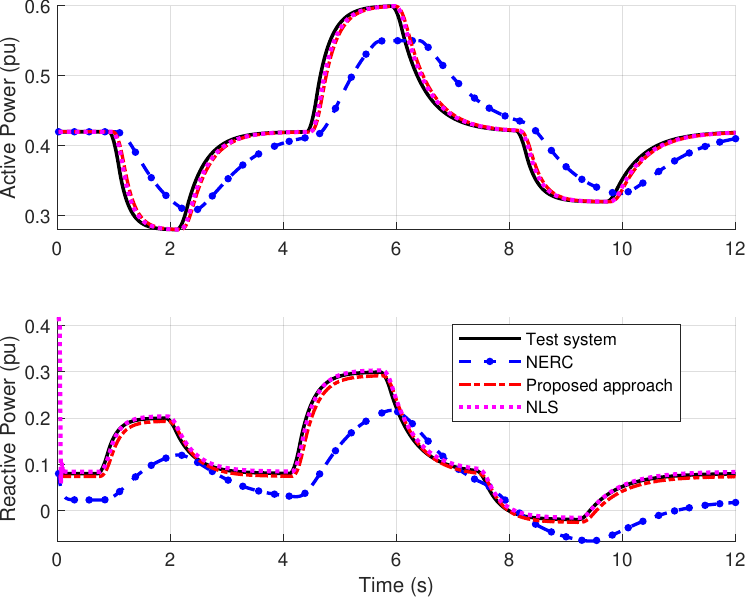}
\caption{Active and reactive power responses under the composite voltage profile.}
\label{fig:event2}
\end{figure}

{\color{black}
To compare DER\_A model responses obtained using NERC guideline \cite{NERC2026} and our calibrated parameters, trajectory error metrics are evaluated for active and reactive power. Specifically, we use the normalized root-mean-square error (NRMSE) and goodness-of-fit (GoF), which are defined as follows:
\begin{align}
\mathrm{NRMSE}_y &= \frac{\sqrt{\frac{1}{N}\sum_{k=1}^{N} e_y(k)^2}}{y_{\text{max}} - y_{\text{min}}}, \\
\mathrm{GoF}_y &= (1 - \mathrm{NRMSE}_y)\times 100\%,
\end{align}
where $e_y(k) = y_{\text{test}}(k) - y_{\text{DER\_A}}(k)$, and $y$ denotes either active or reactive power. As shown in Table~\ref{tab:nerc_vs_calibration_combined}, the normalized error decreases for both active and reactive power across the two cases, with a larger reduction for reactive power. Correspondingly, the goodness-of-fit improves, indicating closer agreement between the DER\_A model and the test system. This behavior is consistent with the voltage-driven disturbance, which primarily excites the voltage--reactive power control dynamics of the aggregated DERs. The resulting improvement in fit reflects a more representative aggregate model of inverter-based resources.

\subsection{Estimation results under a composite voltage disturbance}
To further evaluate the DER\_A model under richer excitation, a composite voltage profile is considered, defined as a sequence of piecewise-constant voltage perturbations with multiple magnitude steps. The resulting profile induces repeated transitions across voltage levels and their associated dynamic responses. For comparison, a nonlinear least-squares (NLS) approach \cite{9637939} is implemented, in which the parameter vector is estimated by minimizing the trajectory mismatch between the measured and simulated active and reactive power responses over the entire time horizon.

Fig.~\ref{fig:event2} shows the active and reactive power responses of the test system compared with the NERC parameter set, the proposed estimation approach, and the NLS-based results under the composite voltage profile. Both the proposed approach and the NLS method closely track the measured response across successive voltage perturbations, accurately capturing both transient behavior and steady-state levels. In contrast, the NERC parameter set exhibits noticeable deviations, particularly during dynamic transitions, indicating reduced accuracy in representing the system response.

Table~\ref{tab:composite_comparison} provides the corresponding quantitative comparison. The proposed approach substantially improves the trajectory fit relative to the NERC guideline, reducing the active- and reactive-power RMSE by 75.75\% and 85.90\%, respectively. The NLS method provides comparable fitting accuracy, but requires 333.42~s and 83 iterations, compared with 5.27~s for the EKF and 27.45~s for the UKF. Note that the NLS approach relies on a batch optimization over the entire data window, making it less suitable for online implementation, unlike the proposed approach. These results indicate that the proposed approach achieves comparable accuracy with significantly lower computational burden and is practical for online deployment.

While we show numerical results under a voltage disturbance, the proposed framework is not restricted to event-based calibration, as parameter estimation depends on the level of excitation in the input--output data. Under ambient conditions, where excitation is limited, the framework remains applicable. This work thus focuses on calibrating DER\_A models from measurement data rather than on constructing the aggregation models themselves. The approach applies in both offline and online operations. The Kalman filtering formulation enables recursive estimation and real-time tracking of parameter variations.
}

\begin{table}[!t]
\centering
\caption{Trajectory error metrics for Cases 1 and 2 (active and reactive power responses)}
\label{tab:nerc_vs_calibration_combined}
\renewcommand{\arraystretch}{1.0}
\setlength{\tabcolsep}{3pt}
\begin{tabular}{llcccc}
\toprule
\textbf{Case} & \textbf{Parameter set}
& $\mathrm{NRMSE}_P$
& $\mathrm{NRMSE}_Q$
& $\mathrm{GoF}_P(\%)$
& $\mathrm{GoF}_Q(\%)$ \\
\midrule
\multirow{2}{*}{1}
& NERC values & 0.1961 & 0.2980 & 86.39 & 83.21 \\
& Estimated values & 0.0746 & 0.0631 & 93.54 & 96.69 \\
\midrule
\multirow{2}{*}{2}
& NERC values & 0.2064 & 0.2487 & 79.36 & 85.13 \\
& Estimated values & 0.0750 & 0.0631 & 92.75 & 95.69 \\
\bottomrule
\end{tabular}
\end{table}

\begin{table}[!t]
\centering
\caption{Quantitative comparison under the composite voltage profile}
\label{tab:composite_comparison}
\renewcommand{\arraystretch}{1.00}
\setlength{\tabcolsep}{3pt}
\begin{tabular}{lccc}
\toprule
\textbf{Method} 
& \textbf{GoF$_\text{P}$ (\%)} 
& \textbf{GoF$_\text{Q}$ (\%)} 
& \textbf{Avg Computation Time (s)} \\
\midrule
NERC & 84.945 & 76.097 & -- \\
Proposed approach & 96.350 & 96.629 
& \begin{tabular}[c]{@{}c@{}} 
5.27 (EKF) \\ 
27.45 (UKF) 
\end{tabular} \\
NLS & 95.863 & 93.718 & 333.42 \\
\bottomrule
\end{tabular}
\end{table}


\section{Conclusions}\label{sec.V}
This paper develops a Kalman filtering framework to tune the parameters of the DER\_A model. We show that an extended Kalman filter converges to parameter values that yield consistent active and reactive power trajectories, matching those of the reference system. We also validate the framework by using an unscented Kalman filter under the same conditions, demonstrating similar estimation performance and confirming the effectiveness of Kalman filtering for tuning DER\_A parameters. Notably, our study finds that DER\_A dynamics are more observable through power measurements than current measurements and identifies the least observable parameters under the chosen model configurations. These results support the practicality of using Kalman filtering methods for utility-scale dynamic modeling of distributed resources, especially when precise tuning is necessary without extensive offline optimization. Future work will explicitly include voltage-tripping logic to represent voltage-dependent disconnection during abnormal conditions and expand the approach to incorporate sensitivity analysis of unmeasured disturbances and other aggregate models, such as those involving battery energy storage. 

\section*{Acknowledgment}
{\color{black}We thank Akhila Kandivalasa for discussions on observability analysis and the anonymous reviewers for their insightful suggestions, which helped improve an earlier version of this paper.}

\appendix
{\color{black}We present the Jacobian corresponding to the augmented state used in the EKF formulation for both cases. Recall that}
\begin{equation}
\bm{F}_{k-1} = \bm{I} + dt \frac{1}{6}(\bm{J}_1 + 2\bm{J}_2 + 2\bm{J}_3 + \bm{J}_4),
\end{equation}

\begin{figure*}[!hb]
\noindent\makebox[\linewidth]{\rule{\textwidth}{0.4pt}}
\begin{align}
\bm{J}_{1}&=
\left[
\arraycolsep=6.5pt\def\arraystretch{0.67}
\begin{array}{c c c c c c c c c c}
-\frac{1}{T_{\text{rv}}} & \cdot & \cdot & \cdot & \cdot & \frac{x_1 - V}{T_{\text{rv}}^2} & \cdot & \cdot & \cdot & \cdot \\
\frac{Q_{\text{ref}}}{T_{\text{iq}} \cdot V_f^2} & -\frac{1}{T_{\text{iq}}} & \cdot & \cdot & \cdot & \cdot & \cdot & \cdot & \frac{x_2 - \frac{Q_{\text{ref}}}{V_f}}{T_{\text{iq}}^2} & \cdot \\
\frac{A_1}{T_{\text{g}}} & \frac{A_2}{T_{\text{g}}} & -\frac{1}{T_{\text{g}}} & \cdot & \cdot & \cdot &\frac{A_3}{T_{\text{g}}} & \frac{A_4}{T_{\text{g}}} & \cdot & \cdot \\
\cdot & \cdot & \cdot & -\frac{1}{T_{\text{pord}}} & \cdot & \cdot & \cdot & \cdot & \cdot & \frac{x_4 - P_{\text{ref}}}{T_{\text{pord}}^2} \\
\cdot & \cdot & \cdot & \cdot & -\frac{1}{T_{\text{g}}} & \cdot & \cdot & \cdot & \cdot & \cdot \\
\cdot & \cdot & \cdot & \cdot & \cdot & \cdot & \cdot & \cdot & \cdot & \cdot \\
\cdot & \cdot & \cdot & \cdot & \cdot & \cdot & \cdot & \cdot & \cdot & \cdot \\
\cdot & \cdot & \cdot & \cdot & \cdot & \cdot & \cdot & \cdot & \cdot & \cdot \\
\cdot & \cdot & \cdot & \cdot & \cdot & \cdot & \cdot & \cdot & \cdot & \cdot \\
\cdot & \cdot & \cdot & \cdot & \cdot & \cdot & \cdot & \cdot & \cdot & \cdot
\end{array}
\right] \quad\textnormal{for case 1}, \label{eq.J1.case1} \\
\bm{J}_{1}&=
\arraycolsep=6.5pt\def\arraystretch{0.67}
\left[\begin{array}{c c c c c c c c c c c c c c}
-\frac{1}{T_{\text{rv}}} & \cdot & \cdot & \cdot & \cdot & \cdot & \cdot & \cdot & \cdot & \cdot & \cdot & \cdot & \cdot & \cdot \\
\cdot & -\frac{1}{T_{\text{p}}} & \cdot & \cdot & \cdot & \cdot & D_1 & \cdot & D_2 & \cdot & \cdot & \cdot & \cdot & \cdot \\
D_3 & D_4 & -\frac{1}{T_{\text{iq}}} & \cdot & \cdot & \cdot & \cdot & \cdot & \cdot & \cdot & \cdot & \cdot & \cdot & \cdot \\
\cdot & \cdot & H_1 & -\frac{1}{T_{\text{g}}} & \cdot & \cdot & \cdot & \cdot & \cdot & \cdot & \cdot & \cdot & \cdot & \cdot \\
\cdot & \cdot & \cdot & \cdot & -\frac{1}{T_{\text{rf}}} & \cdot & \cdot & \cdot & \cdot & \cdot & \cdot & D_5 & \cdot & \cdot \\
D_6 & \cdot & \cdot & \cdot & D_7 & D_8 & D_9 & D_{10} & \cdot & \cdot & \cdot & \cdot & D_{11} & D_{12} \\
\cdot & \cdot & \cdot & \cdot & \cdot & \cdot & D_{13} & \cdot & \cdot & \cdot & \cdot & \cdot & \cdot & \cdot \\
D_{14} & \cdot & \cdot & \cdot & \cdot & \cdot & D_{15} & -\frac{1}{T_{\text{g}}} & \cdot & \cdot & \cdot & \cdot & \cdot & \cdot
\end{array}\right] \quad\textnormal{for case 2}. \label{eq.J1.case2}
\end{align}
\end{figure*}

\noindent
where each \(\bm{J}_i = \frac{\partial \bm{k}_i}{\partial \bm{x}^{\prime}}\) is the Jacobian of the increment \(\bm{k}_i\) with respect to the augmented state vector. We obtain \(\bm{J}_{1}\)\(\in \mathbb{R}^{10 \times 10} \) evaluated at state vector \( \bm{x}_k^{\prime} \) as in \eqref{eq.J1.case1}, where
\begin{align*}
{A}_1 &= \frac{\partial \text{SSF}(\psi)}{\partial \psi} \left[ -\frac{\partial \text{SSF}(I_\text{qv})}{\partial I_\text{qv}} \frac{\partial I_\text{qv}}{\partial x_1} \right] \\
{A}_2 &= \frac{\partial \text{SSF}(\psi)}{\partial \psi}\\
{A}_3 &= \frac{\partial \text{SSF}(\psi)}{\partial \psi} \left( -\frac{\partial \text{SSF}(I_\text{qv})}{\partial I_\text{qv}} \frac{\partial I_\text{qv}}{\partial k_{\text{qv}}} \right) \\
{A}_4 &= \frac{\partial \text{SSF}(\psi)}{\partial \psi} \left( -\frac{\partial \text{SSF}(I_\text{qv})}{\partial I_\text{qv}} \ \frac{\partial I_\text{qv}}{\partial \text{dbd2}} \right) \\
\psi &= x_2 - \text{SSF}(I_\text{qv}).
\end{align*}

Similarly for the case 2, the Jacobian matrix \(\bm{J}_1 \in \mathbb{R}^{14 \times 14}\) is obtained as in \eqref{eq.J1.case2}, where
\begin{align*}
D_1 &= \frac{1}{T_{\text{p}}} \ \frac{\partial \text{SSF}(x_7)}{\partial x_7} \\
D_2 &= \frac{1}{T_{\text{p}}^2} \ \left( x_2 - \text{SSF}(x_7) \right) \\
D_3 &= -\frac{\tan^{-1}\left( \frac{Q_{\text{ref}}}{P_{\text{ref}}} \right) x_2 \frac{\partial V}{\partial x_1}}{T_{\text{iq}}\, V^2} \\
D_4 &= \frac{\tan^{-1}\left( \frac{Q_{\text{ref}}}{P_{\text{ref}}} \right)}{T_{\text{iq}} \, V} \\
{H_1} &= \frac{\partial \text{SSF}(\psi)}{\partial \psi} \ \left( -\frac{\partial \text{SSF}(I_\text{qv})}{\partial I_\text{qv}} \ \frac{\partial I_\text{qv}}{\partial x_3} \right)
\end{align*}
\begin{align*}
\psi &= x_3 - \text{SSF}(I_\text{qv}) \\
D_5 &= \frac{1}{T_{\text{rf}}^2} \ (x_5 - \text{freq}) \\
u_D &= 
\left[ D_{\text{dn}}\text{SDBF}(x_5 - \text{freq}) \right]^{0}_{-\infty}
+ 
\left[ D_{\text{up}}\text{SDBF}(x_5 - \text{freq}) \right]^{\infty}_{0} \\
A &= P_{\text{ref}} + u_D - x_2 \\
B &= x_6 + k_{\text{pg}} \ \text{SSF}(A) \\
D_6 &= -k_{\text{ig}} \ \frac{\partial \text{SSF}(A)}{\partial A} \\
&\quad - k_{\text{w}} \ \frac{\partial \text{SSF}(B)}{\partial B} \ (-k_{\text{pg}}) \ \frac{\partial \text{SSF}(A)}{\partial A} \\
D_7 &= -D_{\text{dn}}\frac{\partial\text{SSF}(D_{\text{dn}}\text{SDBF}(x_5 - \text{freq}))}{\partial(x_5 - \text{freq})} \frac{\partial \text{SDBF}(x_5 - \text{freq})}{\partial x_5} \\
&\quad -D_{\text{up}}\frac{\partial \text{SSF}(D_{\text{up}} \text{SDBF}(x_5 - \text{freq}))}{\partial (x_5 -\text{freq})} \frac{\partial \text{SDBF}(x_5 - \text{freq})}{\partial x_5} \\
D_8 &= -k_{\text{w}} \frac{\partial \text{SSF}(B)}{\partial {B}} \\
D_9 &= k_{\text{w}} \frac{\partial \text{SSF}(B)}{\partial B} \, {A} \\
D_{10} &= \frac{\partial \text{SSF}(A)}{\partial A} \\
D_{11} &= \frac{\partial \text{SSF}(D_{\text{dn}} \ \text{SDBF}(x_5 - \text{freq}))}{\partial (x_5 - \text{freq})} \ \text{SDBF}(x_5 - \text{freq}) \\
D_{12} &= \frac{\partial \text{SSF}(D_{\text{up}}\text{SDBF}(x_5 - \text{freq}))}{\partial (x_5 - \text{freq})}\text{SDBF}(x_5 -\text{freq})\\
D_{13} &= -\frac{1}{T_{\text{pord}}} \ \frac{\partial \text{SSF}(x_7)}{\partial x_7} - k_{\text{w}} \\
D_{14} &= \frac{1}{T_{\text{g}}} \ \frac{\partial \text{SSF}(x_8)}{\partial x_8} \ \left( -\frac{1}{V^2} \right) \\
D_{15} &= \frac{1}{T_{\text{g}}} \ \frac{\partial \text{SSF}(x_7)}{\partial x_7}.
\end{align*}

\noindent
All other rows and columns are zero. The remaining matrices $\{\bm{J}_2,\, \bm{J}_3,\, \bm{J}_4\}$ are computed respectively for both cases.

\bibliographystyle{IEEEtran}
\bibliography{lib}

\end{document}